\documentclass{aa} 

\usepackage[utf8]{inputenc}
\usepackage[english]{babel}
\usepackage[varg]{txfonts}
\usepackage{natbib}
\usepackage{algorithm}
\usepackage{algorithmic}
\usepackage[normalem]{ulem}

\usepackage[caption=false,font=footnotesize]{subfig}
\usepackage[pdftex,
            pdfauthor={Konstantinos Themelis},
            pdftitle={Weak lensing mass reconstruction using sparsity and a Gaussian random field},
            pdfcreator={pdflatex}]{hyperref}
            
\DeclareFontFamily{OT1}{pzc}{}
\DeclareFontShape{OT1}{pzc}{m}{it}{<-> s * [1.10] pzcmi7t}{}
\DeclareMathAlphabet{\mathpzc}{OT1}{pzc}{m}{it}
\DeclareMathOperator*{\argmin}{arg\,min}
\DeclareMathOperator*{\argmax}{arg\,max}

\DeclareMathOperator*{\kw} {\boldsymbol{\kappa}_G}
\DeclareMathOperator*{\kg} {\boldsymbol{\kappa}_G}
\DeclareMathOperator*{\ks} {\boldsymbol{\kappa}_{NG}}
\DeclareMathOperator*{\gb} {\boldsymbol \gamma}
\DeclareMathOperator*{\kb}{\boldsymbol \kappa}
\DeclareMathOperator*{\cov} {\boldsymbol \Sigma}
\DeclareMathOperator*{\covn} {\boldsymbol {\Sigma_n}}
\DeclareMathOperator*{\covk} {\boldsymbol {\Sigma_\kappa}}

\DeclareMathOperator*{\A}{ \mathbf{A}}
\DeclareMathOperator*{\At}{ \mathbf{{\A}^{*}}}
\DeclareMathOperator*{\ifft} { \mathpzc{F}^{*}}

\DeclareMathOperator*{\dft} { \boldsymbol{F}}
\DeclareMathOperator*{\idft} { \boldsymbol{F}^{*}}

\usepackage{color}
\usepackage{fontawesome} % Little github symbol

\newcommand{\nblink}[1]{\href{github_repo_link/folder_with_plot_notebooks/blob/master/#1.ipynb}{\faFileCodeO}}
\newcommand{\github}{\href{https://github.com/CosmoStat/cosmostat}{\faGithub}}

\title{Weak lensing mass reconstruction using sparsity and a Gaussian random field}
\author{J.-L. Starck \inst{\ref{inst1}}
	    \and K.~E. Themelis \inst{\ref{inst1}}
	    \and N. Jeffrey \inst{\ref{inst2},\ref{inst3}}
	    \and A. Peel \inst{\ref{inst4}}
	    \and F. Lanusse \inst{\ref{inst1}}}

\institute{AIM, CEA, CNRS, Universit\'e Paris-Saclay, Universit\'e de Paris, F-91191 Gif-sur-Yvette, France \label{inst1} 
\and Laboratoire de Physique de l'Ecole Normale Sup\'erieure, ENS, Universit\'e PSL, CNRS, Sorbonne Universit\'e, Universit\'e de Paris,\\
Paris, France\label{inst2}
\and Department of Physics \& Astronomy, University College London, Gower Street, London, WC1E 6BT, UK\label{inst3}
\and Institute of Physics, Laboratory of Astrophysics, Ecole Polytechnique F\'ed\'erale de Lausanne (EPFL), Observatoire de Sauverny, 1290 Versoix, Switzerland \label{inst4}}

\abstract{}{We introduce a novel approach to reconstruct dark matter mass maps from weak gravitational lensing measurements. The cornerstone of the proposed method lies in a new modelling of the matter density field in the Universe as a mixture of two components: (1) a sparsity-based component that captures the non-Gaussian structure of the field, such as peaks or halos at different spatial scales; and (2) a Gaussian random field, which is known to well represent the linear characteristics  of the field. 
}{We propose an algorithm called MCALens which jointly estimates these two components. MCAlens is based on an alternating minimization incorporating both sparse recovery and a proximal iterative Wiener filtering. 
}{Experimental results on simulated data show that the proposed method exhibits improved estimation accuracy compared to state-of-the-art mass map reconstruction methods. \github}{}


\begin{document}
\maketitle

% \tableofcontents

% \noindent \NJ{Is it possible to generate all the image plots with the same colour scheme? At the moment there are at least four being used. I recommend a perceptually-uniform colour map ({\url{https://matplotlib.org/3.3.1/tutorials/colors/colormaps.html}}) which would also have the advantage of being colour-blind interpretable. \\ The ``jet'' colour map is particularly egregious ({\url{https://jakevdp.github.io/blog/2014/10/16/how-bad-is-your-colormap/}})}

\section{Introduction}
\label{sec:intro}
In recent years, there has been increasing interest in exploring the two most dominant components of the universe, namely dark matter and dark energy. To this end, large-scale imaging and spectroscopic surveys are currently under development, such as the Euclid mission \citep{laureijs2011euclid}, the Rubin Observatory Legacy Survey of Space and Time \citep{abell2009lsst}, and the Roman Space Telescope \citep{spergel2015wide}, that will map the sky with unprecedented accuracy. A prominent cosmological probe for these surveys is weak gravitational lensing. %, which is sensitive to both visible and invisible matter.

% A prominent cosmological probe is weak gravitational lensing,  and 
% in utilising weak lensing measurements in order to explore the distribution of dark matter and further improve on the constraints of our cosmological models.  
% Weak lensing is thus a powerful tool to probe the mass distribution of various structural components of the universe, including non-visible dark matter
% Weak gravitational lensing, which , provides a powerful tool to probe the mass of large-scale structures in the Universe. 
% with the goal to increase the mapped volume of the Universe In the weak lensing regime, 
% According to the weak gravitational lensing formalism, the distortion of images of distant galaxies is caused by two phenomena generally expressed in two lensing phenomena

Weak gravitational lensing measures correlations in the small distortions of distant galaxies caused by the gravitational potential of massive structures along the line of sight.
Its impact on distant source galaxies is twofold: the galaxy shapes are magnified by a convergence field, $\kappa$, while the galaxies' ellipticities are perturbed from their underlying intrinsic value by a shear field, $\gamma$. 
To contribute constraints on cosmological parameters and models, two-point correlation functions of the shear have been utilised with considerable success \citep{kilbinger13,alsing16}.
This type of correlation is sufficient to statistically describe the Gaussian structures of the lensing field, such as those expected to be present either in the early universe or in large scales that are less affected by gravitational collapse. 
To capture non-Gaussian structures such as those expected in smaller scales at later time, higher order moments of the shear need to be employed \citep{munshi17}.

On the other hand, the convergence density field is not observed directly as a result of the mass-sheet degeneracy \citep{bartelmann01,kilbinger15}. 
Physically, the convergence field reveals the projected total matter density along the line of sight, weighted by a lensing kernel in the mid-distance between the observer and the galaxy sources. The density field is inhomogeneous, encompassing Gaussian-type large-scale structures, as well as non-Gaussian features, such as peaks. To shed light on the way the convergence density field constrains cosmology, peak statistics \citep[e.g.][]{jain00,marian11,lin15,liu16,peel17,fluri18,li19,ajani2020} and higher order correlation functions and moments, such as the Minkowksi functionals \citep[e.g.][]{kratochvil12,shirasaki12,petri13}, have been applied directly on mass maps. It is therefore essential for mass mapping methods to preserve both Gaussian and non-Gaussian features during the reconstruction process.

% To shed light on the way the convergence density field affects galaxy sizes, 
% mass maps have been constructed for many surveys. 
% To extract information from the convergence field, . 
% Such applications of the mass maps,  
% mass mapping methods have been proposed to recover the underlying matter distribution in the universe from the shear measurements, \cite{kaiser93, schneider96, vikram15, oguri17, jeffrey18} 
% It outperforms significantly other methods for peaks recovery.
% It is therefore essential to understand the workings of mass map reconstruction algorithms. This 
Mass mapping methods solve an ill-posed problem due to the irregular sampling of the lensing field and the low signal-to-noise ratio on small scales. A widely used algorithm to perform mass mapping is the Kaiser-Squires method \citep{kaiser93}, which is expressed as a simple linear operator in Fourier space. However, it is inevitable for this estimator to suffer from poor results, since it does not take special care of the noise or missing data. A different approach, motivated also by the Bayesian framework, is that of Wiener filtering \citep{wiener49}. In this approach a Gaussian random field is assumed as a prior for the convergence map, which is responsible for inserting some bias that prevents our estimate from over-fitting \citep{zaroubi95}. Moreover, a recently proposed state-of-the-art method is the Gravitational Lensing Inversion and MaPping using Sparse Estimators (GLIMPSE2D) algorithm \citep{lanusse16}. GLIMPSE2D is a highly sophisticated algorithm that takes advantage of the sparse regularisation framework to solve the ill-posed linear inverse problem. GLIMPSE2D is based on sparse representations (i.e. wavelets), and is therefore well designed to recover piece-wise smooth features. 
An analytical comparison between these three estimators is provided in \citet{starck:jeffrey18}. 
%%%%  JLS TO ADD:  Unfortunately, both these approaches may not be sufficient to actually model the real data. Specifically, the Gaussian prior considered in the Wiener filtering approach can only capture the latent low-frequency part of the signal and has a typical smoothing effect on the recovered field. On the other hand, the sparsity-enforcing $\ell_1$ norm regularization approach of \cite{lanusse16} may as well capture the high-frequency components of the data, such as peaks, but it is unable to model the smoothly varying signal components. In the following , we show how a two mixtures modelling leads to get the best of these two approaches. 

In this paper we propose to bridge the gap between the sparse regularisation method of GLIMPSE2D and the Wiener filtering method by modelling the matter density field in the universe using both linear and non-linear characteristics. Specifically, we assume that the density field is modelled as a mixture of two terms: (1) a non-Gaussian term that adopts a sparse representation in a selected wavelet basis \citep{starck:book15}, 
and (2) a Gaussian term that is modelled using a Gaussian random field \citep{elsner13,Horowitz2019}. The non-Gaussian signal component is able to capture the non-linear characteristics of the convergence field, such as peaks, while the Gaussian component of the signal is responsible for capturing the lower-frequency characteristics of the underlying field, such as smooth variations. To our knowledge, this is the first time that this mixture modelling is proposed for mass map reconstruction. % To solve the proposed optimization task we formulate a two-step optimization process. First, we utilize the GLIMPSE2D algorithm to estimate the non-Gaussian component of the density, whereas, in the second step, we employ a Wiener filtering approach on the shear residuals in order to estimate the Gaussian signal component. It should be noted that for implementing Wiener filtering we employ a proximal calculus method recently presented in \citep{bobin2012cmb}. Experimental results on simulated shear data provide evidence that the proposed method improves the estimation performance of the GLIMPSE2D algorithm in mass mapping recovery. 

This paper is structured as follows. In Sect. \ref{sec:weak_lensing} we introduce the formalism of weak gravitational lensing and describe the mass map reconstruction problem. To this end, we provide a brief overview of the state-of-the-art algorithms of Kaiser-Squires, Wiener filtering, and GLIMPSE2D. We then present our proposed mass mapping method in Sect. \ref{sec:proposed}. The method is novel in the sense that it exploits both sparsity in the wavelet domain as well as a Gaussian random field model. Section \ref{sec:exper} illustrates the enhanced estimation performance of the proposed method by providing experiments conducted on both simulated and real data. 

\emph{Notation}: We use $(\cdot)^*$ to denote the Hermitian transpose of a matrix or the adjoint operator of a transform. With $\left\| \cdot \right\|_1$ and $\left\| \cdot \right\|_2$ we denote the $\ell_1$  and $\ell_2$ norm respectively, $(\left\|\mathbf{x}\right\|_1 = \sum_{i=1}^{N}|x_i|,\  \left\| \mathbf{x} \right\|_2^2=\mathbf{x}^T\mathbf{x})$. The determinant of a matrix or the absolute value of a scalar is denoted by $\left| \cdot \right|$, while $\mathrm{diag}({\mathbf x})$ stands for a diagonal matrix, that contains the elements of vector ${\mathbf x}$ on its diagonal. Finally, $\mathcal{R}^N$ is the $N$-dimensional Euclidean space, $\mathbf{0}$ denotes the zero vector, $\mathbf{1}$ the all-ones vector, and $\mathbf{I}_K$ is the $K \times K$ identity matrix.

\section{Weak lensing mass mapping }

\label{sec:weak_lensing}
Gravitational lensing describes the phenomenon where the light emitted from distant galaxies is deflected as it passes through a foreground mass distribution. The lensing effect causes the images of distant galaxies to be distorted, with the distortion being proportional to the size and shape of the projected matter distribution along the line of sight. 
Specifically, the mapping between the source coordinates, $\beta$, and the lensed image coordinates, $\theta$, is given by the lens equation \citep[e.g.][]{kilbinger15},
\begin{align}
 \boldsymbol \beta = \boldsymbol \theta - \nabla \psi(\boldsymbol \theta),
 \label{eq:coordinates}
\end{align}
where $\psi(\cdot)$ defines the lensing potential that conceals the deflection of light rays by the gravitational lens. 
Under the Born approximation, which assumes that the lensing potential is \emph{weak} enough, we may linearize the coordinate transformation of Eq.~(\ref{eq:coordinates}) by utilising the Jacobian $\mathpzc{A} = \partial \boldsymbol \beta / \partial \boldsymbol \theta$ as,
% its first order Taylor expansion. Then, the linear mapping is expressed via 
% and define the amplification matrix as the Jacobian $\mathpzc{A} = \partial \boldsymbol \beta / \partial \boldsymbol \theta$.
% We can also characterize the lensing transformation utilising the 
% Using the first order Taylor expansion of (\ref{eq:coordinates}), to linearize 
% Assuming that , we may utilize the Born approximation to characterize the lensing effect on the galaxies using the Jacobian of the transformation $\mathpzc{A}_{ij} = \partial \beta_i / \partial \theta_j $. ccc
\begin{align}
 \beta_i = \mathpzc{A}_{ij} \theta_j,
 \label{eq:coordinatesII}
\end{align}
where $\mathpzc{A}_{ij} = \partial \beta_i / \partial \theta_j$ are the coefficients of the amplification matrix $\mathpzc{A}$, and we assume the Einstein summation convention. The symmetrical matrix $\mathpzc{A}$ can also be parameterised in terms of the \emph{convergence}, $\kappa$, and the \emph{shear}, $\gamma$, as, 
\begin{align}
 \mathpzc{A} = \left[\begin{array}{cc}1 - \kappa - \gamma_1 & -\gamma_2 \\  -\gamma_2  & 1 - \kappa + \gamma_1 \end{array} \right].
 \label{eq:amp_matrix}
\end{align}
The convergence can then be defined as a dimensionless quantity that relates to the lensing potential through the Poisson equation, 
\begin{align}
\kappa = \frac{1}{2} (\partial_1\partial_1 + \partial_2\partial_2) \psi = \frac{1}{2} \nabla^2 \psi, 
 \label{eq:poissonI}
\end{align}
while the shear is mathematically expressed as a complex field, whose components also relate to $\psi$, 
\begin{align}
 \gamma_1 = \frac{1}{2}(\partial_1\partial_1 - \partial_2\partial_2)\psi \ \mathrm{and} \ \gamma_2 = \partial_1 \partial_2 \psi.
\label{eq:poissonII}
\end{align}
From Eq.~(\ref{eq:amp_matrix}), we see that the convergence causes an isotropic change in the size of the source image, since it appears in the diagonal of $\mathpzc{A}$. In comparison, the shear causes anisotropic changes to the image shapes. 
The convergence $\kappa$ can also be interpreted via Eq.~(\ref{eq:poissonI}) as a weighted projection of the mass density field between the observation and the source. 
Factoring out the term $(1 - \kappa)$ in Eq.~(\ref{eq:amp_matrix}) leaves the amplification matrix dependent on the reduced shear,
\begin{align}
 \mathpzc{A} = (1-\kappa) \left[\begin{array}{cc} 1 - g_1 & - g_2 \\  - g_2  & 1+ g_1 \end{array} \right ],
 \label{eq:amp_matrixII}
\end{align}
which is directly measured in lensing surveys and it is defined as $g = \gamma / (1 - \kappa )$. 
In the weak lensing limit, where $ \gamma, \kappa \ll 1$, the reduced shear is approximately equal to the true shear, i.e., $g \simeq \gamma$. 

In this paper we are interested in recovering the convergence $\kappa$ from the reduced shear data. This is an ill-posed inverse  problem due to the finite sampling of the reduced shear over a limited area of the survey and the presence of shape noise in the measurements. In the following we review some of the state-of-the-art weak lensing mass reconstructing algorithms, namely the Kaiser-Squires, the Wiener filtering, and the GLIMPSE2D methods.

{\bf \emph{Kaiser-Squires:}}
A theoretical framework for reconstructing convergence maps from the observable weak lensing shear in the Fourier domain was proposed in \citet{kaiser93}. As we have seen in Eqs.~(\ref{eq:poissonI}) and (\ref{eq:poissonII}), the convergence and shear are both expressed as second order derivatives of the lensing potential. Their interrelation via the lensing potential $\psi$ is expressed via a two-dimensional convolution \citep{kaiser93}, 
\begin{align}
 \boldsymbol \gamma (\boldsymbol \theta) = \frac{1}{\pi} \int_{\mathbb{R}^2} \mathrm{d}^2 \boldsymbol \theta'\, \mathpzc{D}( \boldsymbol \theta - \boldsymbol \theta') \kb(\boldsymbol \theta'),
 \label{eq:shear_conv}
\end{align}
where $\mathpzc{D}( \boldsymbol \theta ) = -1 / (\theta_1 - i \theta_2)^2$. This convolution is equivalently expressed in Fourier space as the element-wise multiplication, 
\begin{align}
\tilde{\boldsymbol \gamma}(\boldsymbol k) =  \pi^{-1} \tilde{\mathpzc{D}}(\boldsymbol k) \tilde{\kappa}(\boldsymbol k), 
 \label{eq:shear_conv_F}
\end{align}
where $\boldsymbol k$ is the wavevector, and the Fourier transform of the kernel $\mathpzc{D}( \boldsymbol \theta )$ is given by,
\begin{align}
\tilde{\mathpzc{D}}( \boldsymbol k ) = \pi \frac{ k_1^2 - k_2^2 + 2 i k_1 k_2 }{ k_1^2 + k_2^2},
 \label{eq:kernel_F}
\end{align}
with $k_1$ and $k_2$ being the two frequency components of $\boldsymbol k$. 
Discretising Eq.~\ref{eq:shear_conv} and adopting matrix notation, we may consider that the observed shear $\boldsymbol \gamma$ is generated as a linear combination of the convolution matrix $\mathbf{A}$ and the unknown underlying field $\kb $, i.e., 
\begin{align}
\boldsymbol \gamma = \mathbf{A} \kb + \mathbf{n},
\label{eq:model}
\end{align}
where $\mathbf{n}$ is the statistical uncertainty vector associated with the data. 
% which is assumed to be zero-mean Gaussian, $\mathbf{n} \sim \mathcal{N}(\mathbf{0},\boldsymbol \Sigma_{n})$, where $\boldsymbol \Sigma_{n}$ is the noise covariance matrix.
% The matrix $\mathbf{A}$ represents a set of known linear operations that are responsible for the transformation from convergence to shear in the weak lensing limit. 
Based on Eq.~\ref{eq:shear_conv_F}, $\mathbf{A}$ can be decomposed in Fourier space as 
$\mathbf{A} = \dft \mathbf{P} \idft $, where $\dft$ denotes the discrete Fourier transform,
 $\idft$ is its adjoint, and $\mathbf{P}$ is the diagonal operator that defines the convergence field-shear relation in Fourier space, namely
\begin{align}
 \tilde{\boldsymbol \gamma} = \mathbf{P} \tilde{\kb} = \left( \frac{\mathit{k}_1^2 - \mathit{k}_2^2}{k^2} + \imath \frac{2 \mathit{k}_1 \mathit{k}_2}{\mathit{k}^2}\right) \tilde{\kb}
 \label{eq:shear_conv_DF}
\end{align}
% \FL{And there is no $\pi$ here anymore...} \NJ{The $\pi$ cancels in 8 and 9. I think it is all consistent and matches the Bartlemann \& S review - I deleted the other comments after talking to JL}
where $\mathit{k}^2 = \mathit{k}_1^2 + \mathit{k}_2^2$ and $\tilde{\kb} = \dft \kb$.

% The objective of this paper is to solve the inverse problem of estimating the convergence field $\boldsymbol \kappa$ after observing the noisy shear measurements $\boldsymbol \gamma$. 
% The widely used Kaiser-Squires reconstruction, \cite{kaiser93}, considers the pseudo-inverse of matrix $\mathbf{A}$, which is defined as, ??? %$\mathbf{A}^\dag$.
% \begin{align}
%  [\mathbf{A}^\dag]_{ij} = \frac{k_{1,i}^2-k_{2,i}^2 - 2 k_{1,i} k_{2,i}}{k_{1,i}^2 + k_{2,i}^2}\delta_{i,j},
%  \label{eq:A_pseud}
% \end{align}
% where $\delta_{i,j}$ is a function dependence e from to . 

Equation (\ref{eq:shear_conv_DF}) corresponds to a discretised version of Eq.~(\ref{eq:shear_conv_F}). As it stands, the Kaiser-Squires inversion of Eq.~\ref{eq:shear_conv_DF} suffers from several drawbacks. First, it is not defined for $\boldsymbol{k}=0$, which stems from the mass-sheet degeneracy (i.e. the mean value of the convergence field cannot be retrieved). Next, it is ill-posed, because typically the shear field is a discrete under-sampling of the underlying convergence field. Also, it does not take into account masked data. Nonetheless, the Kaiser-Squires estimate is still used in practice due to its simplicity. 

%~ Therefore, the mean value of $\boldsymbol \kappa$ cannot be recovered from the shear maps. This is a special instance of the well known mass-sheet degeneracy in the weak lensing reconstruction if only shear information is available

{\bf \emph{ Wiener filtering:}} The Wiener filter was introduced in the $1940$s \citep{wiener49}, and it is the optimal linear filter in the minimum mean square sense that provides a denoised version of the desired signal. The Wiener filtered estimate of the convergence map can be expressed using the linear equation
\begin{align}
\kw = \mathbf{W} \boldsymbol \gamma,
\label{eq:thery_wiener1}
\end{align}
where the matrix $\mathbf{W}$ is given by,
\begin{align}
\mathbf{W} = (\A \covk \At+ \covn )^{-1}   \At  \covk,
\label{eq:theory_wiener2}
\end{align}
and $\cov_{\kappa} $ and $\cov_n$ are respectively the pre-defined covariance matrix of the convergence and the noise.
When the noise is not stationary, the Wiener solution is not straightforward, and an iterative approach is required. This will be discussed in  the next section.
% Recently published state-of-the-art methods address this problem by considering utilizes a Gaussian-type prior over $\boldsymbol \kappa$, e.g., \cite{zaroubi95, jeffrey18}

 From a Bayesian perspective, Wiener filtering is equivalent to the maximum a posteriori  (MAP) estimator given that the convergence is a zero-mean Gaussian signal with covariance $\cov_\kappa$. Indeed, assuming that  shear measurements are distorted by uncorrelated Gaussian noise, the likelihood function  shares the noise properties, that is, 
\begin{align}
p(\gb | \kb, \covn) = |2 \pi \covn |^{-\frac{1}{2}} \mathrm{exp} \left [ -\frac{1}{2} (\gb  - \A \kb )^* {\covn}^{-1} (\boldsymbol \gamma  - \A  \kb )\right ],
 \label{eq:likelihood}
\end{align}
while the distribution of the Gaussian random field can be written as
\begin{align}
 p(\kb| \covk) = |2 \pi \covk|^{-\frac{1}{2}} \mathrm{exp}\left[ -\frac{1}{2} {\kb}^* \covk \kb \right].
 \label{eq:kappa_prior}
\end{align}
Given Eqs.~(\ref{eq:likelihood}) and (\ref{eq:kappa_prior}) above, the MAP estimator is expressed as
\begin{align}
\kw & = \argmax_{\kb} \left \{ p(\kb| \boldsymbol \gamma) \right \}  \propto \argmax_{\kb} \left \{ p(\boldsymbol \gamma | \kb, \covn) p(\kb | \covk) \right \} 
\nonumber \\ 
& \propto \argmax_{\kb} \mathrm{exp}\left[ -\frac{1}{2} {\kb}^* \covk \kb -\frac{1}{2} (\gb  - \A \kb )^*{ \covn}^{-1} (\gb - \A \kb ) \right] ,
% & \propto \argmax_{\boldsymbol \kappa} \mathrm{exp}\left[ -\frac{1}{2} ( \boldsymbol \kappa - \mathbf{W}  \boldsymbol \gamma)^* ( \cov_\kappa^{-1} + \mathbf{A} \cov_n^{-1} \mathbf{A}^*)  ( \boldsymbol \kappa - \mathbf{W}  \boldsymbol \gamma) \right],
 \label{eq:map_est}
\end{align}
leading to minimize
\begin{align}
%%  \underset{\boldsymbol \kappa_{G}}{\mathrm{min}} 
\kw & = \argmin_{\kb } \left\{  \| \boldsymbol \gamma  - \A \kb  \|^2_{\covn} + \| \kb \|^2_{\covk} \right\}.
 \label{eq:optw}
\end{align}
It is easy to see that the minimum is attained at $\kw = \mathbf{W} \boldsymbol \gamma $, where  $\mathbf{W}$ is the Wiener filter given in Eq.~(\ref{eq:thery_wiener1}).

The implementation of the Wiener filter requires the inversion of the matrix $\mathbf{W}$ that includes both a signal covariance matrix component and the noise covariance matrix component. 

The unknown $\kw$ is assumed to be a Gaussian random field with a covariance matrix  diagonal in Fourier space, 
 $\covk =   \idft   \boldsymbol{C_\kappa}  \dft $, where $\boldsymbol{C_\kappa} $ is a  diagonal matrix with diagonal values  equal to the theoretical power spectrum $P_{\kappa}$. 
 % \NJ{I think it is actually signal homogeneity (translation invariance, not Gaussianity) that means the signal covariance is diagonal in harmonic space. One could generate a Gaussian random field with Fourier mode covariance.}
When the noise is stationary, its covariance matrix is also diagonal with diagonal elements equal to the noise power spectrum $P_n$. In this case, the filter solution is obtained in Fourier space by $\tilde{\kw} = \tilde{ \mathbf{W} } 
\tilde{\boldsymbol \gamma}$, where the Wiener filter is $\tilde{\mathbf{W} } = \frac{ P_{\kappa} } {P_{\kappa} + P_n} $.
In practice the noise is generally not stationary and depends on the number of shear measurements in the area related to a given pixel of the shear field. 
Therefore the noise covariance matrix is diagonal in pixel space, not in Fourier space, and the Wiener solution becomes more problematic to derive,
requiring either making a wrong assumption (i.e. that the noise is stationary) or inverting a very large matrix.
% Indeed this  matrix to be inverted is clearly dense and of high dimension, i.e., of the size of the number of pixels in the mass map. 
This renders its inversion computationally complex and prone to numerical errors. 
To circumvent this computationally intensive operation, a Forward-Backward (FB) proximal iterative Wiener filtering  
was proposed in \citet{bobin2012wiener} for Cosmic Microwave Background spherical map denoising, exploiting the property
 that the signal and noise covariance matrices are diagonal in pixel and Fourier space, respectively. 
 Eq.~(\ref{eq:optw}) comprises two separable terms, 
\begin{align}
f_1(\kb) =   \parallel  \gb   - \A \kb  \parallel^2_{\covn} \   \mathrm{and} \ f_2(\kb) = \| \kb \|^2_{\covk}.
 \label{eq:optprox}
\end{align}
Following  Forward-Backward methodology \citep{starck:book15}, we can solve Eq.~(\ref{eq:optprox}) by designing an iterative fixed point algorithm as 
\begin{align}
{\kb}^{k+1} = \mathrm{prox}_{\mu  f_2} (\boldsymbol \kappa_{G}^k + \nabla f_1( \boldsymbol \kappa_{G}^k)) ,
\label{eq:fixed}
\end{align}
which is known to converge when $\mu < 2 / \|\At  {\covn}^{-1} \A \|_2$, \cite{combettes05}. 

Computing the proximal operator in Eq.~(\ref{eq:fixed}), we end up with the following  iterative Wiener filtering algorithm,
\begin{itemize}
\item Forward step:  
\begin{align}
\mathbf{t} = {\kb}^n   + 2 \mu \At {\cov}_{n}^{-1} (\gb - \A  {\kb}^n)
\label{eq:alg1}
\end{align}
\item Backward step: 
\begin{align}
{\kb}^{n+1} = \boldsymbol F^{*} \left( \mathbf{P}_{\kb}  \left( \mathbf{P}_\eta +  \mathbf{P}_{\kb} \right)^{-1} \right) \boldsymbol F \mathbf{t},
 \label{eq:alg2}
 \end{align}
\end{itemize}
where $\mathbf{t}$ is an auxiliary variable,  $\mu = \min (\covn) $,  $\mathbf{P}_\eta = 2 \mu \mathbf{I}$ and  ${\kb}^{0}=0$.
This algorithm is free from matrix inversions, since both $\covn$ used in Eq.~(\ref{eq:alg1})
and $\mathbf{P}_{\kb}$ used in Eq.~(\ref{eq:alg2}) are diagonal matrices.
A similar algorithm, exploiting transformations between pixel and harmonic space, was proposed by \citet{elsner13} using the messenger field framework. Such methods can also be adapted to efficiently sample from the posterior distribution of the unknown mass map \citep{alsing_shear_maps,dfe_sampling}.
The Wiener approach recovers the Gaussian component of the convergence field well, but it is far from optimal at extracting the non-Gaussian information from the data, as peak-like structures are suppressed.
This has motivated the development of sparse recovery methods based on wavelets.
% \textsc{GLIMPSE2D}

% Stating the mass mapping problem using Fourier estimators on an irregularly sampled lensing field is an ill-posed linear inverse problem, which we solve using the sparse regularisation framework

{\bf \emph{Sparse Recovery:}}
It has been shown that sparse recovery using wavelets is a very efficient way to reconstruct convergence maps \citep{wlens:starck06,leonard2012,lanusse16,peel17_a520,price20}.
% The GLIMPSE2D algorithm, \citep{lanusse16}, utilizes the sparse regularization framework in order to reconstruct the convergence from the shear measurements. 
The mass mapping problem is addressed as a general ill-posed problem, which is solved via weighted $\ell_1$-norm regularization in a wavelet-based analysis sparsity model. Sparsity in the Discrete Cosine Transform (DCT) domain was also proposed to fill the missing data area in the convergence map  \citep{starck:pires08}.

The GLIMPSE algorithm avoids any binning or smoothing of the input data that could potentially cause loss of information. The primary cost function is
\begin{align}
% \underset{ \boldsymbol \kappa}{\mathrm{min}}
\argmin_{\kb}
\left\{ \frac{1}{2} \| {\gb - \mathpzc{T} {\mathbf{P}} \ifft   \kb} \|^2_{\covn} + \lambda  \| \boldsymbol \omega \odot \boldsymbol \Phi^* \kb\|_1 + \mathpzc{i}_{\mathbb{R}} (\kb)\right \} \label{eq:cost_glimpse}
\end{align}
where $\mathpzc{T}$ is the nonuniform discrete Fourier transform (NDFT) matrix, $\mathbf{P}$ is defined in Eq.~(\ref{eq:shear_conv_DF}), $\lambda$ is a sparsity regularization parameter, $\boldsymbol \omega$ is a weighting vector, $\boldsymbol \Phi^*$ is the adjoint operator of the wavelet transform, and $\mathpzc{i}_{\mathbb{R}}$ is an identity function that drives the imaginary part of the convergence to zero. This cost function is also generalised in GLIMPSE2D in order to replace the shear by the reduced shear, or to incorporate the flexion information.  
It has been shown that GLIMPSE2D significantly outperforms Wiener filtering for peaks recovery, while Wiener filtering does better on the Gaussian map content \citep{starck:jeffrey18}.

\emph{Deep Learning:} 
Deep learning techniques have recently been proposed and seem very promising \citep{starck:Jeffrey2020}.  
The input of the neural network is not the shear field directly, but rather the Wiener solution. We could imagine that the closer we are to the true solution with standard techniques such as Wiener or sparsity, the better deep learning can improve the solution. Open questions related to deep learning remain to be answered, such as its generalization to cosmologies not present in the training data set, or the potential bias introduced by using a theoretical power spectrum in the Wiener solution serving as input of the neural network.

% In this paper we supplement the sparse modeling of GLIMPSE2D with a Gaussian random field in an effort to capture the low scale modes of the convergence as well. 

% Recently published state-of-the-art methods address this problem by considering either a Wiener filtering approach that utilizes a Gaussian-type prior over $\kb$, e.g., \cite{zaroubi95, jeffrey18}, or by exploiting a sparse representation of the convergence field in the wavelet domain, e.g., \cite{lanusse16}. 

% ==========================================================================

% ==========================================================================

\section{Modelling with sparsity and a Gaussian random field}

\label{sec:proposed}

\subsection{A new convergence map model}
We have seen in Sect.~\ref{sec:weak_lensing} two different models: the first modelling the convergence map as a 
 Gaussian random field, leading to good recovery of the large scales of the convergence map but surpression peak structures, and the second assuming the convergence map is compressible in the wavelet domain (i.e. sparse modelling).
The sparse recovery is clearly complementary to the Wiener solution, since it recovers peaks extremely well but the Gaussian content poorly.

To address these limitations, it seems  natural to introduce 
a novel modelling approach, where the convergence field $\kb$ is assumed to comprise two parts, a Gaussian and a non-Gaussian:
\begin{align}
\kb  =  \kg + \ks.
\label{eq:newx}
\end{align}
The non-Gaussian part of the signal $\boldsymbol \kappa_{\mathrm{NG}}$ is subject to a sparse decomposition in a wavelet dictionary, 
while the component $\boldsymbol \kappa_{\mathrm{G}} $ is assumed to be inherently non-sparse and Gaussian. 

The Morphological Component Analysis (MCA) was proposed in \citet{starck:sta04,starck:elad05} to separate two components mixed in a single image when these components have different morphological properties. This looks impossible, since we have two unknowns and one equation, but it was 
shown that it is sometimes possible to extract these two components if we can exploit their morphological differences.
This requires having different penalisation functions $\boldsymbol C_\mathrm{G}$ and $\boldsymbol C_\mathrm{NG}$ on each of these two components, and we need 
to minimise
\begin{align}
\min_{\kg, \ks } \left\{    \| \gb  - \A ( \kg + \ks )  \|^2_{\covn}   + {\boldsymbol C_\mathrm{G}} (  \kg )  +   \boldsymbol C_\mathrm{NG}(   \ks)      \right\}.
\end{align}
    %             % + \| \kb \|^2_{\covk} \end{align}
MCA performs an alternating minimization scheme:
\begin{itemize}
\item Estimate $\kg$ assuming $\ks$ is known:
\begin{align}
\min_{\kg} \left\{    \|  (\gb  - \A  \ks)   - A \kg )  \|^2_{\covn}   + {\boldsymbol C_\mathrm{G}} (  \kg )  \right\}.
\label{eq_kg}
\end{align}
 \item Estimate $\ks$ assuming $\kg$ is known:
\begin{align}
\min_{\ks} \left\{    \|  (\gb  - \A  \kg)   - A \ks )  \|^2_{\covn}   + {\boldsymbol C_\mathrm{NG}} (  \ks )  \right\}.
\end{align}
\end{itemize}

Examples of such decompositions can be seen on the MCA web page\footnote{\url{http://jstarck.cosmostat.org/mca}}. A range of MCA applications in astrophysics can be found in \cite{starck:sta02_3,andre2010,moller2015,bobin2016,melchior018,joseph2019,hgmca2020}.
 
 \subsubsection*{The Gaussian component $\kg$ }
We use here the standard Wiener modeling where $\kg$ is assumed to be a Gaussian random field:
 \begin{align}
 {\boldsymbol C_{G}} (  \kg )   =  \| \kb \|^2_{\covk},
 \end{align}
and the solution of Eq.~(\ref{eq_kg}) is obtained using the iterative Wiener filtering presented in the previous section.

 \subsubsection*{The non-Gaussian component}
 There are different ways to use a sparse model in the MCA framework. The most obvious  would be to use standard  $\ell_1$ or  $\ell_0$-norm regularisation in a wavelet-based sparsity model, as it is done in the GLIMPSE2D algorithm. This would give:
 \begin{align}
 \label{mode_ng}
{\boldsymbol C_{NG}} (  \ks )   =  \lambda  \| \boldsymbol  \Phi^{*} \ks  \|_p,
\end{align}
 where $p=0$ or $1$, $\Phi$ is the wavelet matrix, and $\lambda$ is {the regularization parameter (Lagrange multiplier)}.
 
 After implementing this approach, we found that large wavelet scales and  Fourier low frequencies are relatively close, leading to difficulties in separating the information.
We have therefore investigated another approach, which involves first estimating the set $\Omega$ of active coefficients---i.e. the scales and positions where wavelet coefficients are above a given threshold---typically between 3 and 5 times the noise standard deviation relative to each wavelet
 coefficient.  $\Omega$ can therefore be seen as a mask in the wavelet domain, where $\Omega_{j,x} = 1$ if a wavelet coefficient detected 
 at scale $j$ and position $x$, i.e. when  $ \mid    ( \boldsymbol \Phi^{*}  \At \gb)_{j,x}   \mid  >   \lambda \sigma_{j,x}$, and $0$ otherwise.
The noise $\sigma_{j,x}$ at scale $j$ and position $x$ can be determined using noise realizations as in the GLIMPSE algorithm.
 An even faster approach is to detect the significant wavelet coefficients
  on $ \boldsymbol \Phi^{*}  \At  \covn^{-\frac{1}{2}} \gb$ instead of $ \boldsymbol \Phi^{*}  \At \gb$.
 The noise is therefore whitened, as the noise factor $\At  \covn^{-\frac{1}{2}}$ is Gaussian
  with standard deviation equal to unity and with a uniform power spectrum. We implemented both approaches giving similar results, though the second is simpler.

  Once this wavelet mask  $\Omega$ is estimated, we can estimate the non Gaussian component $ \ks$ by
 \begin{align}
\min_{\ks} \left\{    \|   \Omega  \odot  \boldsymbol \Phi^{*}  \left(  \left(  \gb  - \A  \kg)   - A \ks \right)  \right)  \|^2   + {\boldsymbol C_{NG}} (  \ks )  \right\}.
\label{eq_mcasparse}
\end{align}
 with ${\boldsymbol C_{NG}} (  \ks )   =    \mathpzc{i}_{\mathbb{R}} (\boldsymbol \kappa_{NG})$. 

This changes the original formalism since the data fidelity term is now different, but it presents a very interesting advantage.
Once $\Omega$ is fixed, the algorithm is almost linear and only the positivity constraint remains. 
Therefore, we can easily derive a good approximation of the error map, just by propagating noise and relaxing this positivity constraint. 
This will be further discussed in the following.
{Similarly to the GLIMPSE method, a positivity constraint is applied on the non-Gaussian component $\ks$. Peaks in $\kb$ can be on top of voids, and therefore have negative pixel values. As peaks are captured by the  non-Gaussian component, they are positive by construction in $\ks$, but the convergence map $\kb = \kg  + \ks$ can still be negative at peaks positions. {Larger are the non-Gaussianities, more we can expect MCAlens to improve over linear methods such as the Wiener one. }

The prior signal auto-correlation of the Gaussian component is included within the signal covariance term of the Gaussian component. We encode no explicit prior auto-correlation for the non-Gaussian signal and no explicit prior cross-correlation between the Gaussian and non-Gaussian component. Clearly, such correlations exist but including their contribution in the prior in this framework would be extremely difficult theoretically and in practice. However, such correlations will still appear in the final reconstruction, driven by the correlation information in the data. 
}

% ===========================

\subsection{MCAlens Algorithm}

\begin{algorithm}[ht]
\caption{MCAlens algorithm}
\label{algo:mcalens}
\begin{algorithmic}[1]
\STATE{\textbf{Input}: Shear map $\gb_1,\gb_2$, signal and noise covariance $\covk,\covn$, and detection level $\lambda$.}
% \STATE{Initialize $(\x^{(0)},\mathbf{u}^{(0)}),\W^{(0)}=\mathbf{1},\tau>0,\eta>0,\mu\in\left]0,1\right]$}
\STATE\label{algo:line4}{Initialize: ${\ks}^{(0)}={\kg}^{(0)}=\Omega=0$, $\mu = \min (\covn)$ , $\mathbf{P}_\eta = 2 \mu \mathbf{I}$.}
\STATE\label{algo:coef}{Calculate wavelet coefficients: $ \boldsymbol \alpha = \boldsymbol \Phi^{*}  \At  \covn^{-\frac{1}{2}} \gb$.}
\STATE\label{algo:omega}{$ \forall j,x$,  set $\Omega_{j,x} = 1 ~~~ if ~~~  \mid    (\boldsymbol \alpha )_{j,x}   \mid  >   \lambda  $.}
\FOR{$n=0,\ldots,\text{N}_\text{max}-1$}

\STATE{-----------------  Find  $ \ks $ -------------------- }
\STATE{Calculate the shear residual: $\boldsymbol {\gb}_r =   \gb - \A (  \kg^{(n)} + \ks^{(n)}) $.}
\STATE{Calculate the sparse residual: $\boldsymbol{s_{r}} =   \At   \covn^{-\frac{1}{2}}  \boldsymbol {\gb}_r  $.}
\STATE{Calculate the sparse residual in the mask: \\ $\boldsymbol s_{mr}   =  \boldsymbol \Phi \left(  \Omega  \odot  \left(  \boldsymbol  \Phi^{*}    \boldsymbol s_{r}  \right) \right)$.}
\STATE\label{algo:line5}{Get the new sparse component: $\mathbf{S}  = \ks^{(n)} + \boldsymbol s_{mr}$.}
\STATE\label{algo:line6}{Positivity constraint: $ \ks^{(n+1)} =  [ \mathbf{S}  ]_{+} $.}

\STATE{---------------- Find  $ \kg $  ----------------------- }
\STATE{Calculate the shear residual: $\boldsymbol {\gb}_r =   \gb - \A (  \kg^{(n)} +  \ks^{(n+1)}     ) $.}
\STATE{Forward step: $\mathbf{t} = {\kg}^n   + 2 \mu \At {\cov}_{n}^{-1} \boldsymbol {\gb}_r$.}
\STATE{Backward step: $ {\kw}^{n+1} = \boldsymbol F^{*} \left( \mathbf{P}_{\kb}  \left( \mathbf{P}_\eta +  \mathbf{P}_{\kb} \right)^{-1} \right) \boldsymbol F \mathbf{t}$.}

\ENDFOR
\RETURN $ \left( {\kg}^{ (  \text{N}_\text{max})},  \ks^{(\text{N}_\text{max})} \right).$
\end{algorithmic}
\end{algorithm}

We solve the  recovery problem of Eq.~(\ref{eq:newx}) using a two-step optimization procedure. First, 
a gradient descent step to minimise Eq.~(\ref{eq_mcasparse}) to recover the non-Gaussian component  $\ks$. 
This is followed by an iteration of the iterative Wiener filtering to  minimise Eq.~(\ref{eq:optw}).
Details of the method are given in the Algo.~\ref{algo:mcalens}.

{The number of scales $N_s$ used to compute the wavelet transform 
of an $N_x \times N_y$ image is automatically derived 
by $N_s = \mathbf{int}( \log( \min(N_x, N_y)))$. 
The $\lambda$ parameter is a detection level, which was fixed for all our experiments with real and simulated data to $5$,  i.e. $5$ times the noise standard deviation, which is a conservative threshold which gave excellent results. 
}

 % The total number of wavelet scales which are used is derived automatically from the number of pixels (NbrScale = int (  log ( min([Nx,Ny]) ) ), and all wavelet scales are used to estimated to non-gaussian component. The only parameter is the detection level, which was fixed for all our experiments with real and simulated data  to 5,  i.e. 5 times the noise standard deviation, which is a conservative threshold which gave excellent results. In this sense, we can consider MCAlens is parameter free. 

 \subsection{Errors}
 
Much attention has recently been given to the estimation of errors or uncertainties with mass map products.
For linear methods such as Wiener or Kaiser-Squires, it is easy to estimate the standard deviation (or root-mean-square, RMS) per pixel, just by propagating noise realisations using the same reconstruction filters.
The uncertainty per pixel does not, however, give the probability that a clump in a reconstructed image is true or only due to some noise fluctuations. Another approach, closer to certain science cases with maps, estimates the significance of clumps. 
 In \citet{peel17_a520}, Monte Carlo simulations were used to address the significance of clumps. 
 In  \citet{repetti2018}, a hypothesis test called BUQO was proposed to do the same task, requiring the user to define manually a mask around the clump. Similarly, in~\citet{price_structures}, hypothesis tests of Abell-520 cluster structures using Highest Posterior Density Regions were performed.
 
With MCAlens, we can include aspects of both approaches. 

\subsubsection*{RMS and SNR maps}
In Algo.~\ref{algo:mcalens}, there are two steps involving a non-linear operator, first  for the estimation
of  $\Omega$ in line~\ref{algo:omega} and the second in line~\ref{algo:line6} to perform the positivity constraint.
To propagate noise realizations, $\Omega$ has to be set to the one obtained with the data, and this non-linearity step
therefore does not occur,  so only the positivity remains. 
A full linear algorithm could therefore be obtained just 
by removing this positivity constraint during the noise propagation, by replacing  $ \ks^{(n+1)} =  [ \mathbf{S}  ]_{+} $
 in Algo.~\ref{algo:mcalens} line~\ref{algo:line6} by  $ \ks^{(n+1)} =  \mathbf{S}  $. This leads to more noise
 entering the solution, since few pixel values with negative values in $\ks$ are not thresholded, and the derived RMS map is therefore 
 slightly conservative.
Hence, we can  build noise realizations, run the MCAlens algorithm on each realization, and derive the RMS map by taking pixel 
 per pixel the standard deviation of the obtained reconstructed maps.
The SNR map is derived by dividing the absolute value of the reconstructed map by the RMS map.

\subsubsection*{Significance map}
In Algo.~\ref{algo:mcalens} line~\ref{algo:omega}, 
the wavelet mask $\Omega$ is obtained by comparing the wavelet coefficients,  $ \boldsymbol \alpha = \boldsymbol \Phi^{*}  \At  \covn^{-\frac{1}{2}} \gb$,
to the threshold $\lambda \sigma $ where $\sigma=1$ as the noise is whitened with unit variance. 
This corresponds to performing a hypothesis test $H_0$ 
that the wavelet coefficient is due to noise only, 
and if a given wavelet coefficient $ \boldsymbol \alpha_{j,x}$ is such that $  \mid  \boldsymbol \alpha_{j,x}   \mid  >   \lambda$, then the $H_0$ hypothesis is rejected
 and we assume that the coefficient amplitude cannot be explained by noise fluctuations and is therefore due to signal. 
$\lambda$ is therefore directly related to the significance of the wavelet coefficients, and the mask $\Omega$  
indicates which coefficients 
are detected with a given significance level. 
$\Omega_{j,x}$ is binary, and we build the significance map $s$ by
$s_x = \sum_j \Omega_{j,x}$, i.e. a simple coadding of all binary scales.
Such a map could also be used as a way to automatically derive the user mask required in the BUQO method  \citep{repetti2018}.
We present in Sect.~\ref{sec:exper} examples of RMS, SNR and significance maps.

\subsection{Extension to the Sphere}
\label{sect_sphere}
When a wide-field  map needs to be reconstructed, the flat approximation cannot be used anymore, and we have to build a map on the sphere. 
A traditional approach is to decompose the sphere into overlapping patches, assume a flat approximation on each individual patch, 
 reconstruct each patch independently, and finally recombine all patches on the sphere. This solution is certainly good enough to 
recover clumps relative to clusters, i.e. the non-Gaussian component, but certainly not for the Gaussian component which contains
information at low frequencies. In the framework of the DES project, a $1500$ deg$^2$  map has been reconstructed  \citet{des2018_curvedsky},
using HEALPIX pixelisation \citep{pixel:healpix} and a straightforward spherical Kaiser-Squires algorithm consisting
in:
\begin{enumerate}
\item calculating a spin transform of the HEALPIX shear map to get both the E and B spherical harmonic coefficients, 
\item smoothing by a Gaussian both E and B modes in the spherical harmonic domain, 
\item applying an inverse spherical transform independently on each of these two modes to get the two convergence maps 
$\kappa^E$ and $\kappa^B$.
\end{enumerate}
Sparsity in a Bayesian framework \citep{price_sphere2020} and forward fitting in harmonic space \citep{bacon2020}  have also been recently proposed to do spherical mass mapping. 
Using similarly a HEALPIX shear and convergence pixelisation, we can also easily derive an extension of both the iterative-Wiener filtering and the MCAlens algorithm by
\begin{itemize}
    \item replacing the matrix  $\A  = \dft \mathbf{P} \idft $ in Eq.~(\ref{eq:model})  by $\A =  {\bf _{2}Y}   {\bf _{0}Y}^{*}$,
where $ {\bf _{s}Y}$ and $ {\bf _{s}Y}^{*}$  represent the forward and inverse spin-s spherical harmonic transforms respectively,
\item replacing the
  the wavelet decomposition  $\boldsymbol \Phi$ used in Eq.~(\ref{mode_ng}) by the spherical wavelet decomposition \citep{starck:sta05_2}.
\end{itemize} 
 Then the same MCAlens algorithm given in Algo.~\ref{algo:mcalens} can be used to derive a spherical convergence map. An example is given in Sect.~\ref{exp_sphere}.

\subsection{B-mode}
\label{sect_bmode}

We focused earlier on the convergence map (i.e. E-mode). The MCAlens algorithm given in Algo.~\ref{algo:mcalens} remains valid 
if one wants to estimate both E and B mode by adopting the following notation: 

 $\boldsymbol \gamma = \begin{pmatrix} \boldsymbol \gamma_1 \\
									\boldsymbol \gamma_2 \\
\end{pmatrix}, 
\boldsymbol \kb = \begin{pmatrix} \boldsymbol \kb_E \\
									\boldsymbol \kb_B \\
\end{pmatrix}, 
\A = \begin{pmatrix}
   \frac{\mathit{k}_1^2 - \mathit{k}_2^2}{k^2}  &  \frac{2 \mathit{k}_1 \mathit{k}_2}{\mathit{k}^2} \\ 
   \frac{2 \mathit{k}_1 \mathit{k}_2}{\mathit{k}^2}  & -   \frac{\mathit{k}_1^2 - \mathit{k}_2^2}{k^2}
\end{pmatrix},
\boldsymbol \alpha = \begin{pmatrix} \boldsymbol \alpha_E \\
									\boldsymbol \alpha_B 
\end{pmatrix}, 
\mathbf{P}_{\kb} = \begin{pmatrix}\mathbf{P}_{\kb_E} \\
									\mathbf{P}_{\kb_B} \\
\end{pmatrix}, 
\boldsymbol \Phi = \begin{pmatrix} \boldsymbol \Phi_E  \\
						      \boldsymbol \Phi_B 
\end{pmatrix}$, \\
\noindent where the matrix $\boldsymbol \Phi$ consists in applying a sparse decomposition independently on each mode. In practice we use
the same wavelet decomposition for both (i.e. $\boldsymbol \Phi_E = \boldsymbol \Phi_B$).
The delicate point is the Wiener filter to apply to the B-mode at line 15 of the algorithm. 
In theory, $\mathbf{P}_{\kb_B} = 0$, and by construction,
 no Gaussian component can be recovered. Since the B-mode is mainly useful for investigation of systematic errors, 
 we are not interested in recovering the B-mode least square estimator (which is zero), and 
 we find it more useful to process the B-mode similarly to the E-mode. We therefore advocate rather to use $\mathbf{P}_{\kb_B} = \mathbf{P}_{\kb_E}$. 
 This way the fluctuations in the E-mode can properly be  compared to those in the B-mode.  
 
% If we assume that the signal $\boldsymbol \kappa_G$ and the noise $\mathbf{n}$ are uncorrelated and, hence, their power spectra are associated as  
%\begin{align}
%\mathbf{P}_{\boldsymbol \kappa_G} = [ \mathbf{P}_{\boldsymbol \kappa} - \mathbf{P}_{\boldsymbol \kappa_{NG}} ]_+ %- \mathbf{P}_n
%\label{eq:nspec2}
%\end{align}
%where $\mathbf{P}_{\boldsymbol \kappa_{NG}}$  can be easily computed from the data, while $\mathbf{P}_{\boldsymbol \kappa}$ can be known from cosmological models. In this paper we utilised the approach in (\ref{eq:nspec2}) to obtain a good approximation of $\mathbf{P}_{\boldsymbol \kappa_G}$.
%

% In Niall's paper \cite{jeffrey18} wiener filter is applied on the shear to recover $\boldsymbol \gamma_G$ first and then $\boldsymbol \kappa_G$ is recovered by appling the Kaiser-Squires inversion in Fourier space. 
%, while $\boldsymbol \Sigma_\mathbf{n} $ %and $\mathbf{N}_{\boldsymbol \epsilon} = \mathbf{A} \mathbf{N}_{\mathbf{n}} \mathbf{A}^T$ is the covariance matrix of the noise in the shear and convergence domain, respectively. 
% Discuss the diagonal matrices in pixel and Fourier space
% \begin{align}
%  \boldsymbol \kappa_G &= \boldsymbol \Sigma_{\boldsymbol \kappa_G} \mathbf{A}^T \left[ \mathbf{A} \boldsymbol \Sigma_{\boldsymbol \kappa_G} \mathbf{A}^T + \mathbf{N}_\mathbf{n} \right]^{-1} \boldsymbol \gamma_r\\
%  & = \boldsymbol \Sigma_{\boldsymbol \kappa_G} \left[ \boldsymbol \Sigma_{\boldsymbol \kappa_G} + \mathbf{N}_{\boldsymbol \epsilon} \right]^{-1} \mathbf{A}^{-1} \boldsymbol \gamma_r
% \end{align}
% get the right equations

\section{Experimental results}

\label{sec:exper}
\subsection{Toy Model Experiment}

% MICE simulations used in our previous experiment are very noisy (around 3 gal per $arcmin^2$), 
% and we didn't detect anything in wavelet space. 
% So the MCAlens solution was equal to the Wiener solution and there is no gain in this case.  

We use a simulation derived from RAMSES N-body cosmological simulations  \citep{code:teyssier02},
with a $\Lambda$CDM model (see \url{http://www.projet-horizon.fr}), with a pixel size of $0.34'$  x $0.34'$ and a galaxy redshift
of 1.
% instead of  $4.5'$  x $4.5'$. The resulting convergence map covers a field of 2◦x 2◦  
To get a realistic mask and noise behavior, we use the MICE pixel noise covariance derived for the DES project (see Appendix A). 
% As the map resolution is not the same as the one used in MICE, it lead to a signal-noise ratio better by a factor 20  compared to  what is expected in practice in a project like Euclid.
As the pixel resolution is different, it leads to an optimistic realization, but it has the advantage of illustrating well the impact of our MCA model. % \FL{How many galaxies per arcmin2 does that correspond to?}

\begin{figure*}
\vbox{
\hbox{    \centering
        {
        \includegraphics[width=.4\textwidth]{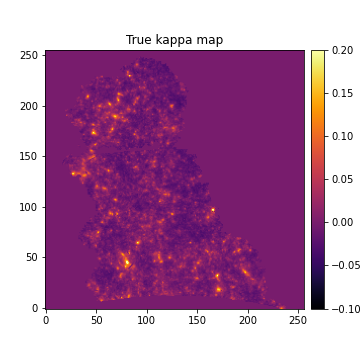}
         \includegraphics[width=.4\textwidth]{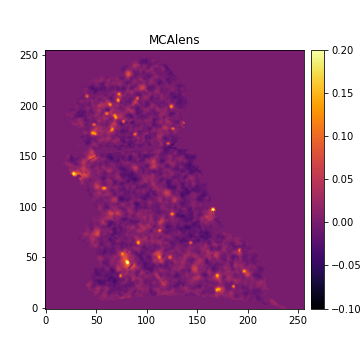}
        }}
\hbox{    \centering
        {
        \includegraphics[width=.4\textwidth]{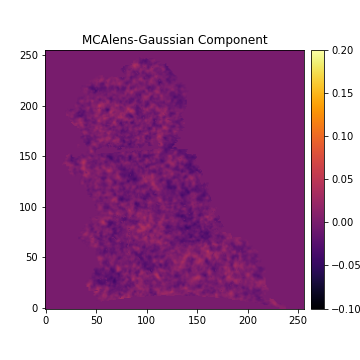}
         \includegraphics[width=.4\textwidth]{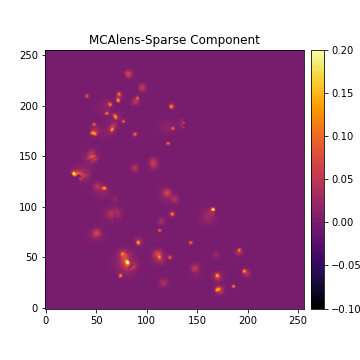}
        }}        
  }
          \caption{RAMSES simulations: Top, true convergence map and MCAlens recovery; bottom, Gaussian components and sparse components. The sum of  these two maps is equal to the top right MCAlens map.}
          \label{fig:ramses_mcalens}
\end{figure*}

\begin{figure*}
\vbox{
\hbox{    \centering
        {
        \includegraphics[width=.3\textwidth]{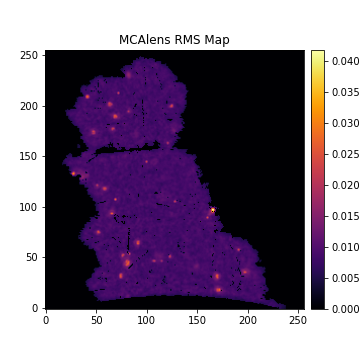}
         \includegraphics[width=.3\textwidth]{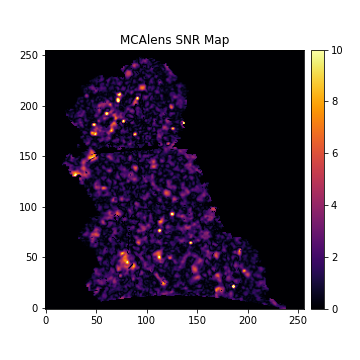}
         \includegraphics[width=.3\textwidth]{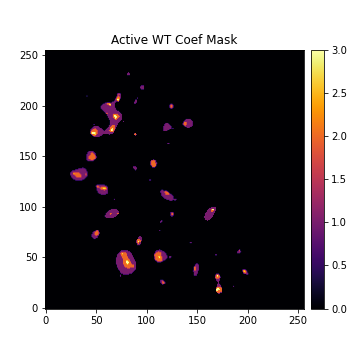}
        }}     
  }
          \caption{RAMSES simulations: from left to right, RMS map, SNR map and significance map.}
          \label{fig:ramses_RMS_mcalens}
\end{figure*}

We have run MCAlens using 100 iterations, $\lambda=5$ (i.e. detection at $5\sigma$), the MICE covariance matrix,
and the used  theoretical power spectrum was the true map power spectrum.
\begin{figure}
        \centering
        {
        \includegraphics[width=.5\textwidth]{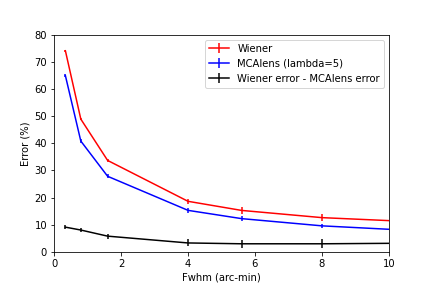}
        \caption{RAMSES simulations:  Error versus scale for Wiener in red and MCAlens in blue.}
        \label{fig:ramses100}
        }
\end{figure}
To evaluate the results, we ran 100 different noise realizations, and we applied both Wiener filtering and MCAlens. We calculated the reconstruction error at different resolutions with  
\begin{align}
Err_{\%} (\sigma)  =  \frac{ \parallel {\boldsymbol G}_\sigma \left( M \left( \kb -  {\kb}_t \right) \right) \parallel }{  \parallel  M  {\kb}_t  \parallel}
\label{eq_err}
\end{align}
where $ \parallel x \parallel = \sqrt{ \sum_i x_i^2}$, $\kb$ is the reconstruct convergence map, 
${\kb}_t$ the true convergence map, 
${\boldsymbol G}_\sigma(x)$ is the convolution of $x$ with a Gaussian with standard deviation $\sigma$,  and $M$ is the binary mask with $M_k = 1$ if the covariance matrix is not infinite at location $k$ (i.e. we have data at this location) and $0$ otherwise.
Figure \ref{fig:ramses100} shows the mean error $Err_{\%} (\sigma)$ for both the Wiener and MCAlens solutions. {The black curve shows the difference between the Wiener error and the MCAlens error, allowing to better visualise that the improvement is larger at fine scales.
It is interesting to note that at MCAlens is better than Wiener even at large scales.
Indeed, when the MCAlens non-Gaussian component contains 
a significant amount of features as in the case of this experiment, these features have  also a non negligible contribution on larger scales, which explain why MCAlens remains better than Wiener at large scales.
}
MCAlens leads to a clear improvement in terms of quadratic error.
The top panels of Fig.~\ref{fig:ramses_mcalens} show respectively the simulated convergence map and the MCAlens reconstructed map. 
The bottoms panels show the Gaussian and the non Gaussian part recovered from the noisy the data.
The sum of these two components is equal to the MCAlens reconstruction (top right).
This shows that both the non-linear  and the linear components can be recovered well. 
Figure \ref{fig:ramses_RMS_mcalens} shows respectively the RMS map, the SNR map and the significance map.

% For comparison, Figures \ref{fig:kappa_glimpse} and \ref{fig:kappa_wiener} show the results from GLIMPSE2D, which recovers only the non-Gaussian features, and the result from Wiener filtering, which recovers only the Gaussian part of the signal.

% From Deriaz et al 2012
% The cosmological model is taken to be in concordance with the $\Lambda$CDM model. We have chosen a model with the following parameters close to WMAP : $\Omega_m = 0.3$, $\sigma_8 = 0.9$, $\Omega_L = 0.7$, $h = 0.7$. Each simulation has $256^3$ particles with a box size of 162 Mpc/h. 
% The resulting convergence map covers a field of $2^{\circ}$ x $2^{\circ}$ with 350 x 350 pixels and assumes a galaxy redshift of 1. 
% The overdensities correspond to the halos of groups and clusters of galaxies.

\subsection{Columbia Lensing Simulation}

We use here a convergence map, released  by the Columbia Lensing group\footnote{\url{http://columbialensing.org}}
\citep{Liu2018MassiveNuSCM}, corresponding to a cosmological model
with  parameters  $ \left\lbrace M_{\nu}, \Omega_m, 10^{9}A_s, M_{\nu}, h, w \right\rbrace $=$ \left\lbrace 0.1, 0.3, 2.1 , 0, 0.7, -1\right\rbrace $,
and with a pixel size of $0.4'$  x $0.4'$. 
%the sum of neutrino masses $M_{\nu}=0$, the total matter density parameter $\Omega_m= 0.3$, the primordial power spectrum amplitude 
%$A_s\cdot 10^{9}=2.1$ at the pivot scale $k_0=0.05$ Mpc$^{-1}$,  the reduced Hubble constant $h=0.7$, the spectral index $n_s=0.97$, the baryon density parameter $\Omega_b = 0.046$ and the dark energy equation of state parameter $w=-1$.
% The fiducial model is set at $ \left\lbrace M_{\nu}, \Omega_m, 10^{9}A_s  \right\rbrace $=$ \left\lbrace 0.1, 0.3, 2.1 \right\rbrace $.
We rebinned the map to  $0.8'$  x $0.8'$,  and similarly to the previous experiment,  
we simulated noisy data using the same MICE covariance,  applying a global rescaling in order to have realistic noise corresponding to 
a mean number of galaxies equal  to 30 per arcmin${}^2$, 
%instead of 3, 
as we expect in future space lensing surveys.
To evaluate the results, we ran 100 different noise realizations, and we applied Kaiser-Squires, sparse recovery, Wiener filtering, and MCAlens. For the sparse recovery and MCAlens, we used $\lambda=5$ (i.e. detection at $5\sigma$), and for Wiener and MCAlens we 
the used the theoretical power spectrum as the true convergence map power spectrum.

\begin{figure}
        \centering
        {
        \includegraphics[width=.5\textwidth]{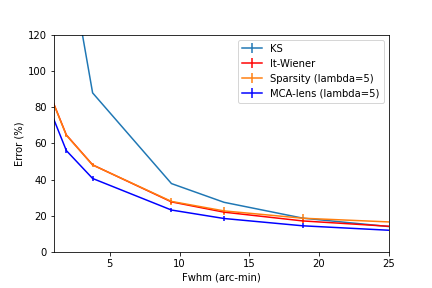}
        \caption{Columbia simulations: error versus scale for four different methods: Kaiser-Squires (light blue), Wiener (red), sparsity (orange) and MCAlens (blue).}
        \label{fig:jia100}
        }
\end{figure}

\begin{figure*}
\vbox{
\hbox{    \centering
        {
        \includegraphics[width=.3\textwidth]{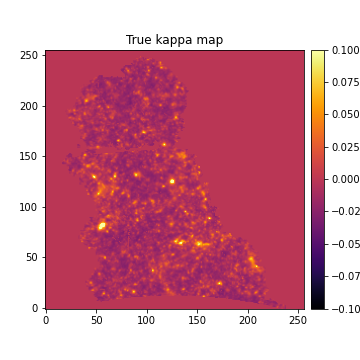}
        \includegraphics[width=.3\textwidth]{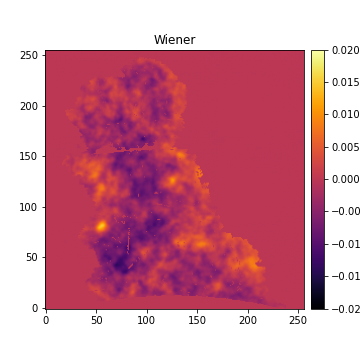}
        \includegraphics[width=.3\textwidth]{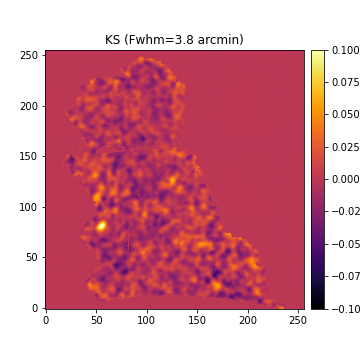}
        }}
\hbox{    \centering
        {
         \includegraphics[width=.3\textwidth]{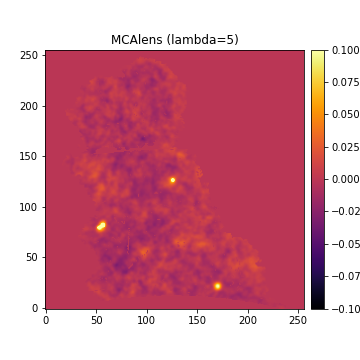}
        \includegraphics[width=.3\textwidth]{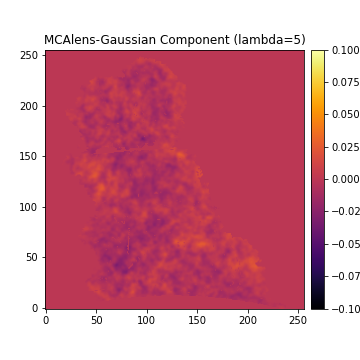}
         \includegraphics[width=.3\textwidth]{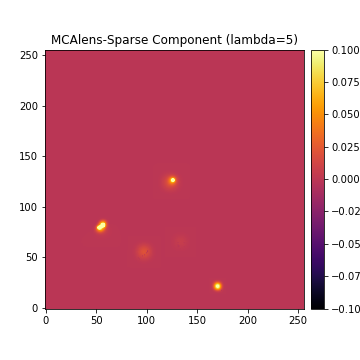}
        }}
\hbox{    \centering
        {         
         \includegraphics[width=.3\textwidth]{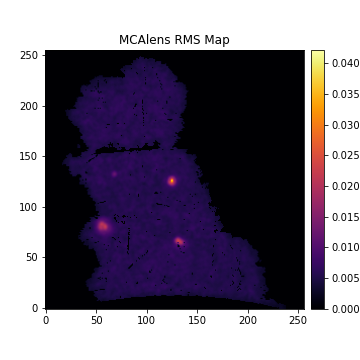}
         \includegraphics[width=.3\textwidth]{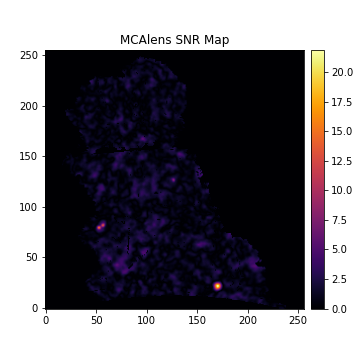}
          \includegraphics[width=.3\textwidth]{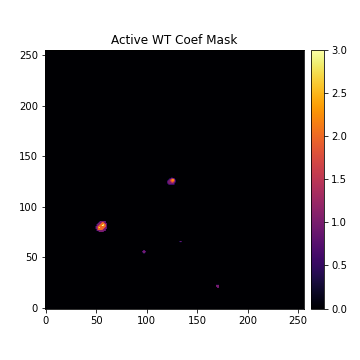}
        }}
 }
\caption{Columbia convergence map recovery:  Top, true convergence map, Wiener map, and Kaiser-Squires map smoothed with a Gaussian having a Full Width at Half Maximum of 3.8 arcmin.
Middle, MCAlens map and its Gaussian and sparse and. The sum of these two last maps is equal to the first one.
Bottom, RMS, SNR and significance maps.}
\label{fig:jia_mcalens}
\end{figure*}

Figure \ref{fig:jia100} shows the error computed using Eq.~(\ref{eq_err}). We can see that, in addition to the well-recovered peaks, MCAlens also leads to a clear improvement in terms of quadratic error compared the other methods. {On the contrary to the RAMSES experiment, we see here a convergence at large scales between MCAlens and Wiener. This is due that the MCAlens non-Gaussian component contains only a few peaks (see Fig.~\ref{fig:jia_mcalens}, middle right), which have therefore a negligible contribution to the largest scales.}

The top row of Fig.~\ref{fig:jia_mcalens} shows respectively the simulated convergence map, the Wiener result, and the Kaiser-Squires reconstruction with smoothing applied. The middle shows the MCAlens map and its Gaussian and non Gaussian parts recovered from the noisy the data.
The sum of these two components is equal to the MCAlens.
Similarly to previous experiments, both the non-linear  and the linear components are well recovered. 
The bottom row of Fig.~\ref{fig:jia_mcalens} shows respectively the RMS map, the SNR map and the  significance map.

%  \STATE{Calculate wavelet coefficients: $ \boldsymbol \alpha = \Phi^{*}  \At  \covn^{-\frac{1}{2}} \gb$.}

%{Repetti}, Audrey and {Pereyra}, Marcelo and {Wiaux}, Yves
% {Scalable Bayesian uncertainty quantification in imaging inverse problems via convex optimization}
%  \citep{repetti2018}

% {Spherical Bayesian mass-mapping with uncertainties: full sky observations on the celestial sphere}
% {Price}, Matthew A. and {McEwen}, Jason D. and {Pratley}, L. and {Kitching}, Thomas D.
% \citep{price2020}

% error bars extract them from code sum the sigmas of the support set per scale 

\subsection{Spherical Data}
% cmd="/Users/starck/git/cosmostat/cosmostat/src/cxx/build/wls_mcalens -F2 -i200 -N covmat.fits -v -s5 -n5  -m4 -S mica_kappa_map_cl.fits mice_g1_noisy.fits mice_g2_noisy.fits res_mice_200"
% /Users/starck/git/cosmostat/cosmostat/src/cxx/build/wls_mcalens -v -m2 -i200  -N covmat.fits -S mica_kappa_map_cl.fits  mice_g1_noisy.fits mice_g2_noisy.fits res_mice_wiener
\label{exp_sphere}

\begin{figure*}
\centerline{
\vbox{
\hbox{
        \includegraphics[width=.35\textwidth]{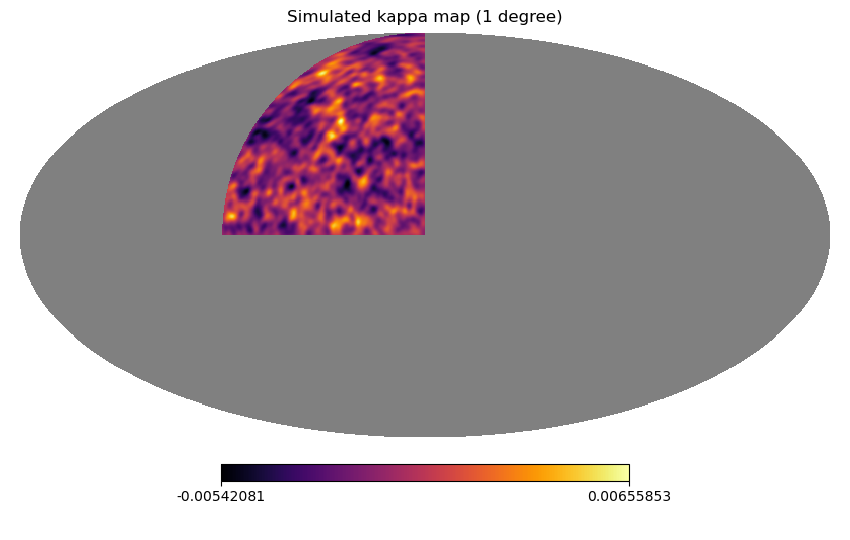}
        \includegraphics[width=.35\textwidth]{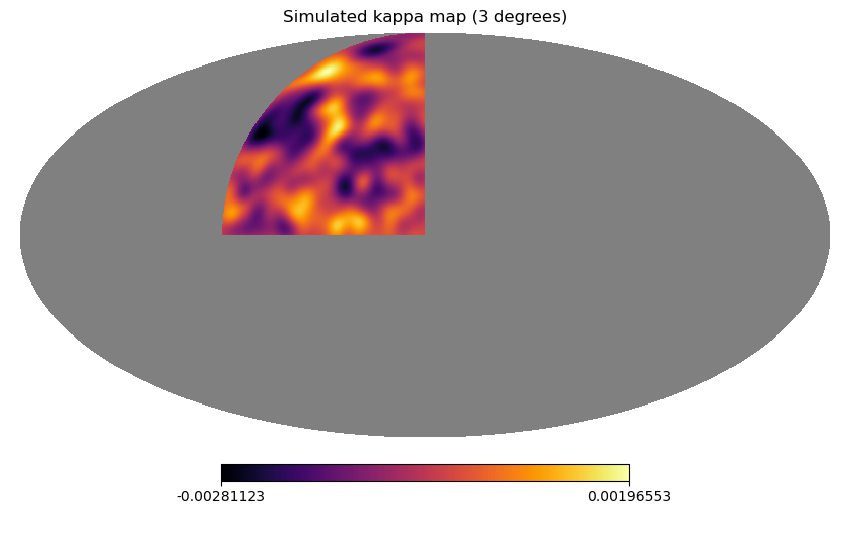}
}
\hbox{
        \includegraphics[width=.35\textwidth]{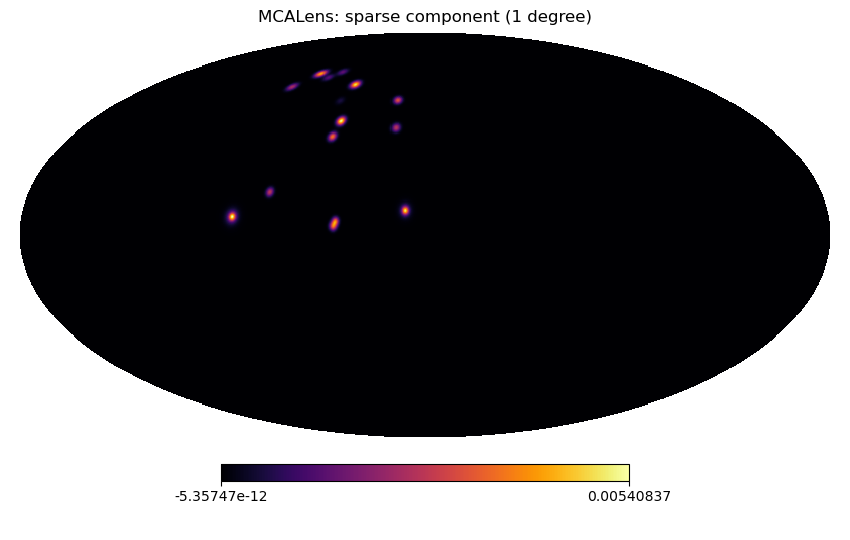}
        \includegraphics[width=.35\textwidth]{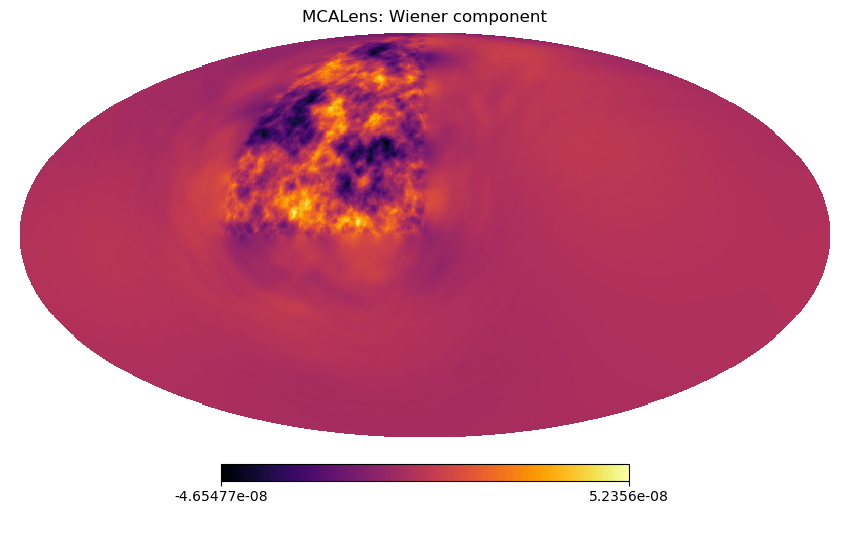}
}}}
        \caption{MICE simulations: top, true convergence map at 1 and 3 degrees. Bottom, MCAlens sparse and Gaussian components.}
        \label{fig:mice_sphere}
\end{figure*}
We have created a full shear map from the  full sky MICE simulated map and its covariance matrix, which has
around 3 galaxies per $arcmin^2$.
We ran  the spherical MCAlens method with  200 iterations, $\lambda=5$ (i.e. detection at $5\sigma$),  
and we used  the power spectrum of the simulation as the theoretical power spectrum.
The top row of Fig.~\ref{fig:mice_sphere} shows the simulated noise-free map at 1 and 3 degrees, while the bottom row shows the MCAlens non-Gaussian and Gaussian components.

\subsection{COSMOS field}

In this last section, we apply MCAlens to reconstruct a convergence map of the 1.64 deg$^2$ HST/ACS COSMOS survey \citep{Scoville2007a}. In this work, we make use of the bright galaxies shape catalogue produced for \citep{Schrabback2010}.

\begin{figure*}
\vbox{
\hbox{    \centering
        {
        \includegraphics[width=.3\textwidth]{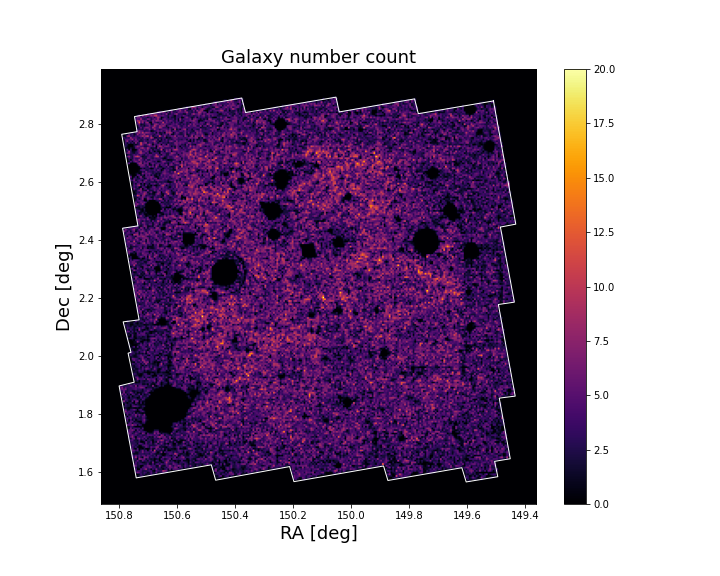}
        \includegraphics[width=.3\textwidth]{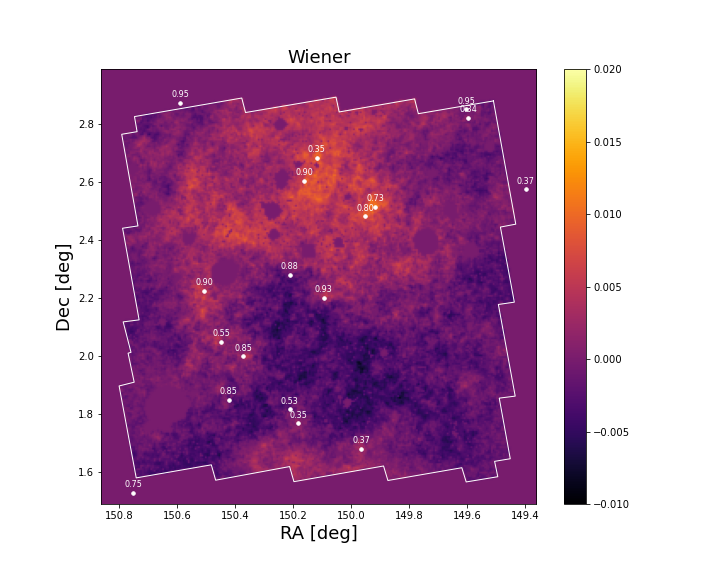}
         \includegraphics[width=.3\textwidth]{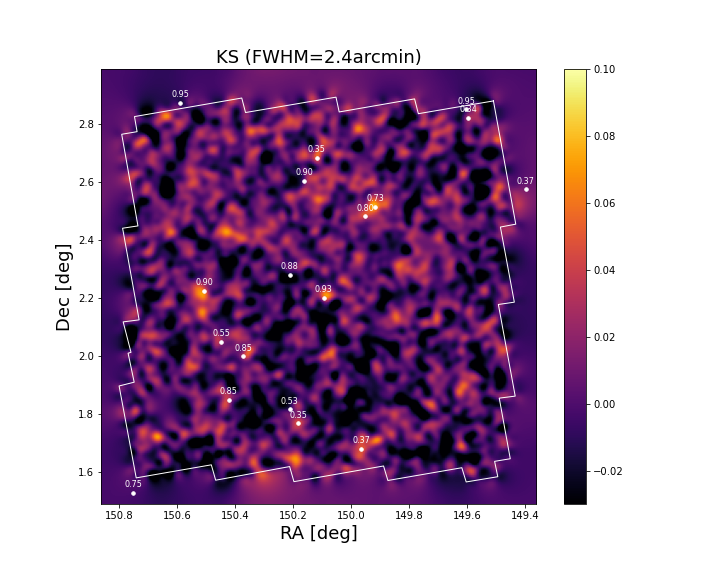}
        }}
\hbox{    \centering
        {
        \includegraphics[width=.3\textwidth]{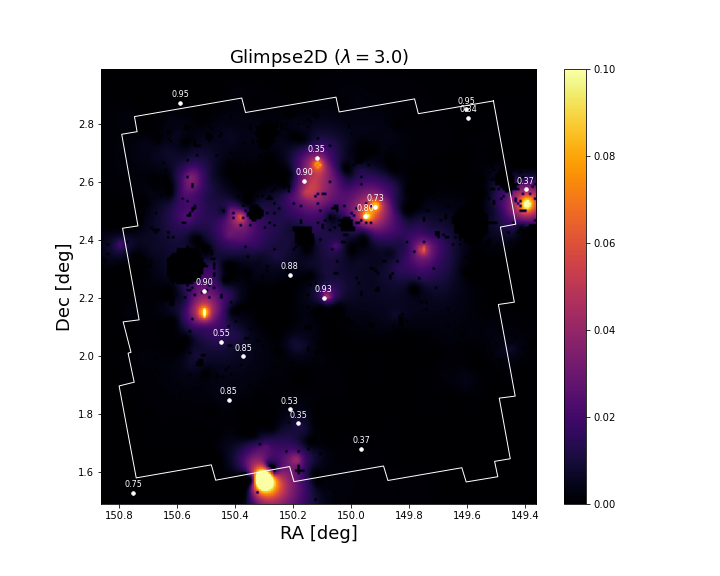}
         \includegraphics[width=.3\textwidth]{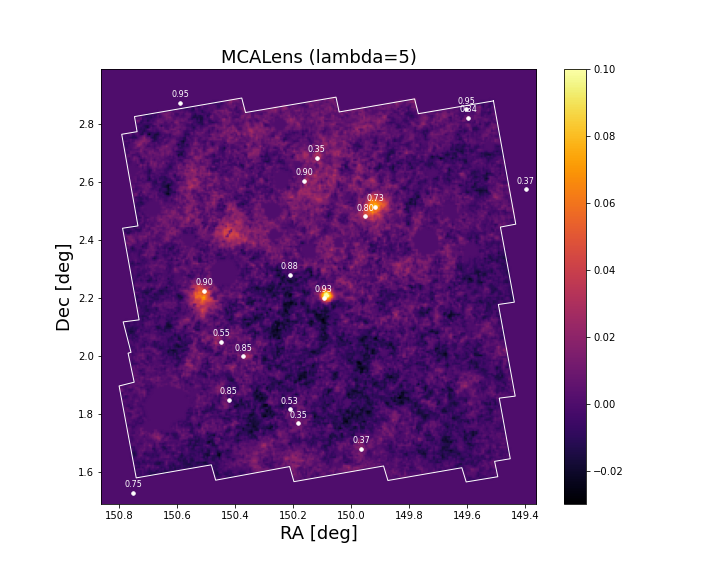}
          \includegraphics[width=.3\textwidth]{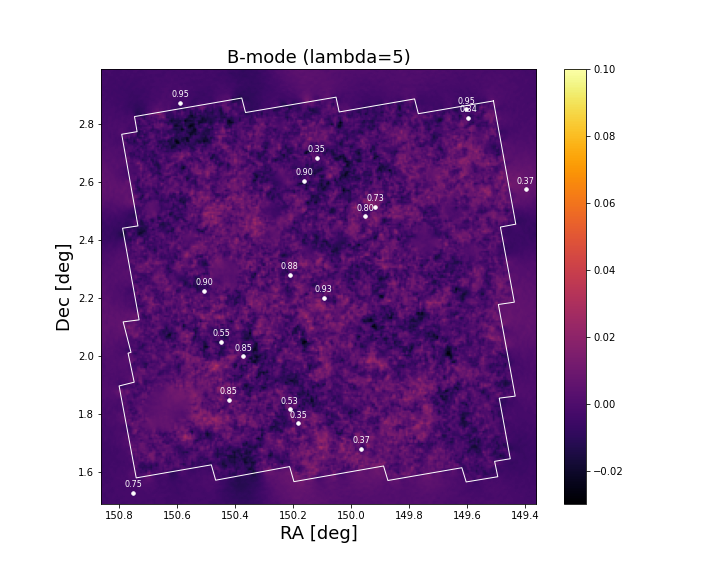}
        }}
%\hbox{    \centering
%        {         
%         \includegraphics[width=.3\textwidth]{FIG/fig_cosmos_db_mcalens_5sigma_xw.png}
%          \includegraphics[width=.3\textwidth]{FIG/fig_cosmos_db_SNR5.png}
%          \includegraphics[width=.3\textwidth]{FIG/fig_cosmos_db_mcalens_5sigma_bmode.png}
%        }}
 }
%\caption{COSMOS data:  Top, galaxies count map, Wiener map, and Massey 2007 map. Middle
%Glimpse, Kaiser-Squires map smoothed with a Gaussian having a Full Width at Half Maximum of 2.4 arcmin, and MCAlens.
%Bottom, MCA Gaussian, SNR map and MCAlens B-mode map.}
\caption{COSMOS data:  Top, galaxies count map, Wiener map, and Kaiser-Squires map smoothed with a Gaussian having a Full Width at Half Maximum of 2.4 arcmin.  
Bottom, Glimpse, MCAlens and MCAlens B-mode map.}
\label{fig:cosmos}
\end{figure*}

The results after applying MCAlens on COSMOS data are presented in Fig.~\ref{fig:cosmos}.
In the top row are the galaxy count map, the Wiener map, and the Kaiser-Squires map smoothed with a Gaussian at a Full Width at Half Maximum of 2.4 arcmin.
The bottom row shows respectively the Glimpse, MCAlens  E-mode and MCAlens B-mode maps. White dots show the locations and redshifts of X-ray selected massive galaxy clusters from the XMM-Newton Wide Field Survey \citep{finoguenov07} with $0.3 < z < 1.0$.

%Fig.~\ref{fig:cosmos}  top the galaxies count map, the Wiener map, and the Massey 2007 map. 
%Fig.~\ref{fig:cosmos}   middle show the Glimpse, the Kaiser-Squires map smoothed with a Gaussian having a Full Width at Half Maximum of 2.4 arcmin, and MCAlens result. Fig.~\ref{fig:cosmos} bottom shows the MCAlens Gaussian component, the SNR map and MCAlens B-mode map.

\section{Conclusion}
A novel mass mapping algorithm has been presented that is able to recover high resolution convergence maps from weak gravitational lensing measurements.  
Our proposed process involves a model with two components, a Gaussian and a non-Gaussian, 
for which we have developed an efficient algorithm to derive the solution.
 We have shown that we can also handle a non-diagonal covariance matrix.
 We have  extended the method so it can deal with spherical maps, which is needed for future surveys such as the Euclid space mission.
 Our experiments clearly show a significant improvement compared to the state of art.
 
 In the spirit of reproducible research, the MCAlens algorithm is publicly available in the CosmoStat's Github package\footnote{\url{https://github.com/CosmoStat/cosmostat}}, including the material needed to reproduce the simulated experiences (folder examples/mcalens\_paper and script make\_fig.py).

% use section* for acknowledgment
\section*{Acknowledgment}
We thank Tim Schrabbaack for sending us his COSMOS shear catalog and  the Columbia Lensing group (\url{http://columbialensing.org}) for making their suite of simulated maps available, 
and NSF for supporting the creation of those maps through grant AST-1210877 and XSEDE allocation AST-140041.

%===================================================================
% \newpage
% \clearpage

\begin{appendix}

\section{MICE simulations}
We use the public MICE (v2) simulated galaxy catalogue, which is constructed from a lightcone N-body dark matter simulation \citep{mice_1,mice_2,mice_3,mice_4,mice_5,cosmohub}. 
The MICE catalogue provides the calculated weak lensing (noise-free) observables: shear and convergence. In a given patch of simulated sky we select galaxies in the redshift\footnote{Redshift due to the expansion of the Universe is used as a proxy for distance to a galaxy, as it is an observable that monotonically increases with distance from an observer.} $z$ range $[0.6,1.4]$. Each galaxy corresponds to a noisy shear measurement, and we subsample with a density of $\sim$8000 galaxies per deg$^2$.

Uncorrelated, complex shape noise values are randomly drawn from a Gaussian distribution and added to the shear value of each selected galaxy. This noise per galaxy is zero mean and has variance $2 \sigma_\epsilon^2 = 0.1636$ (as estimated from data \citep{starck:jeffrey18}).
The final pixelised noise ($\mathbf{n}$) has variance that depends on the number of galaxies per pixel. 

 In our simulated data, we mimic these conditions by choosing to remove all galaxies in given regions. Here there are no shear measurements available and the noise variance is effectively infinite. %  Maps of the noisy input shear data along with the noise variance are shown in Fig. \ref{fig_sim0} with the mask applied. 

\begin{figure*}
\centerline{
\vbox{
\hbox{
        \includegraphics[width=.35\textwidth]{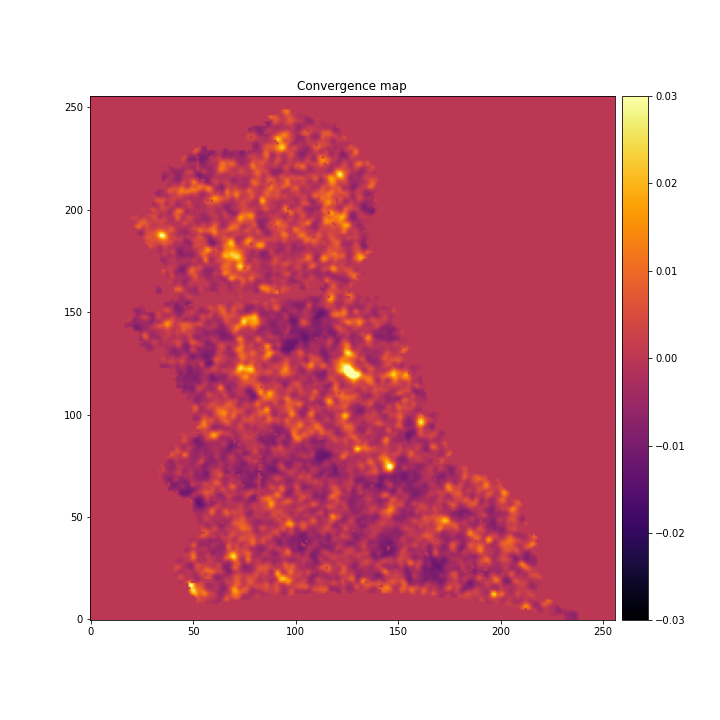}
        \includegraphics[width=.35\textwidth]{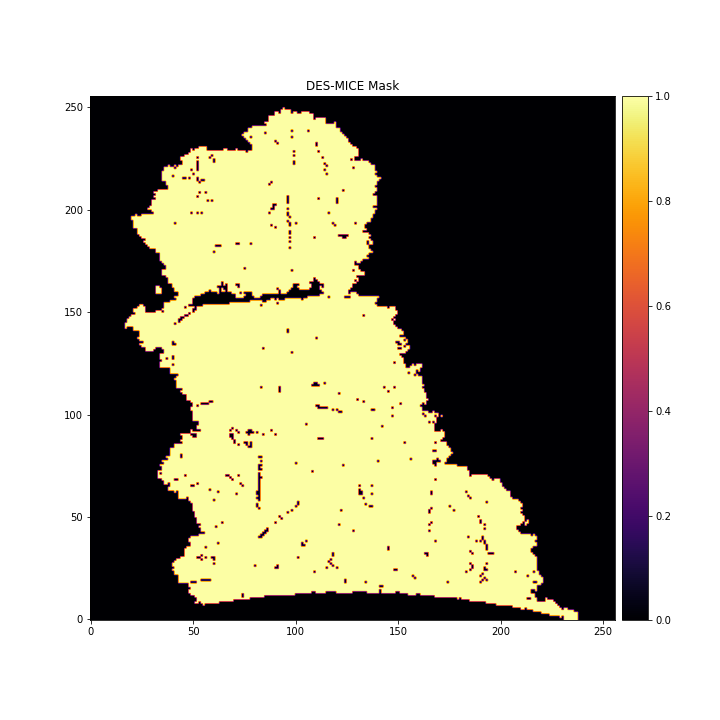}
}
\hbox{
        \includegraphics[width=.35\textwidth]{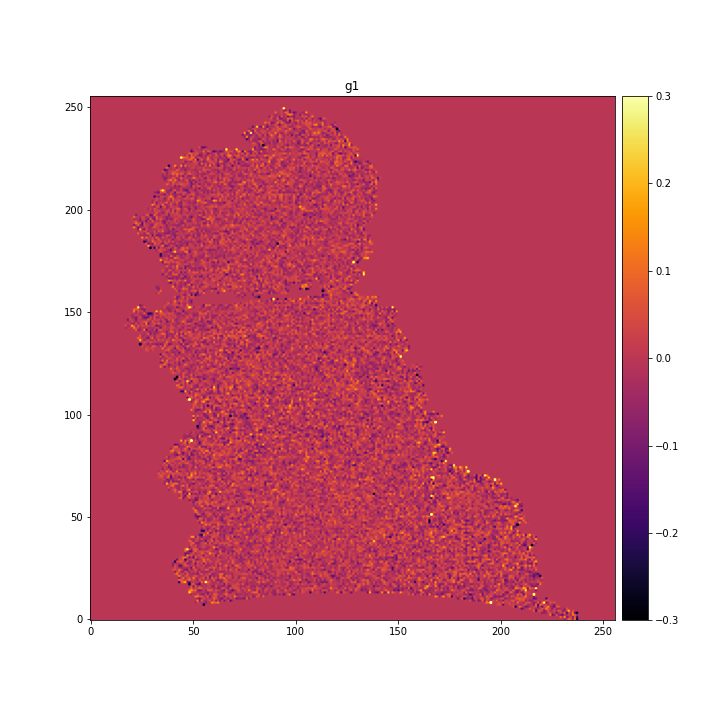}
        \includegraphics[width=.35\textwidth]{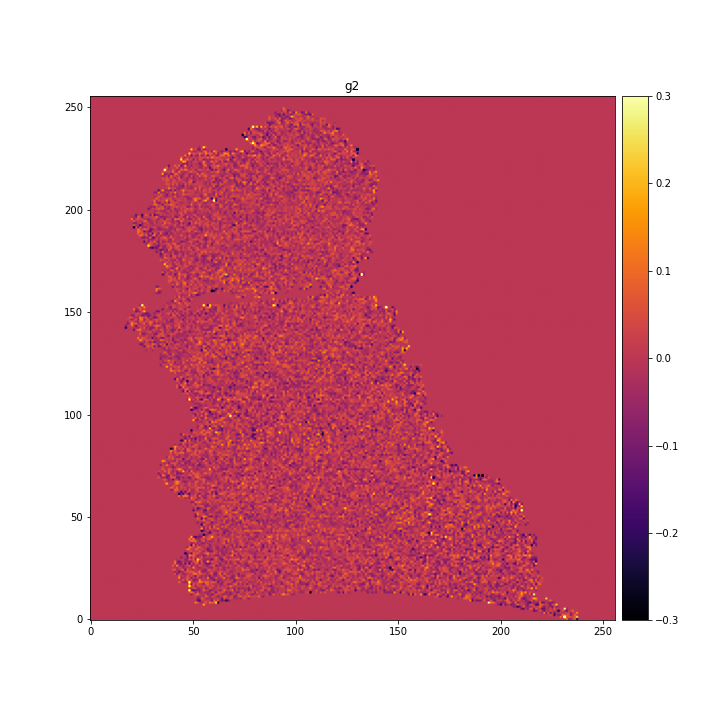}
}}}
        \caption{Top, true MICE convergence map with DES SV footprint and mask. Bottom, Two simulated observed shear components, $\gamma_1$ and $\gamma_2$.}
        \label{fig:mice_shear}
\end{figure*}

The top row of Fig.~\ref{fig:mice_shear} shows a simulated convergence map with the DES SV footprint and mask, and the bottom row shows two simulated observed shear component maps, $\gamma_1$ and $\gamma_2$.

\section{A Wiener tour}
\label{sect_appA}

\subsection{Wiener and inpainting}
Missing data is a common problem for galaxy surveys as foreground objects obscuring the background galaxies have to be ``masked out''.
In addition to the non stationary noise, observed shear fields therefore also present missing data. Noting the mask $M$ as equal to 1 if we have shear measurements at a pixel position and zero otherwise, missing data can be handled in Wiener filtering by forcing the noise covariance matrix to be very high at locations where $M=0$. This leads the solution to be different from zero and smoothed in the missing data area, and remove border effect artefacts. 
The Wiener filtering can therefore be seen as an inpainting technique, since it fills the missing area in the image.
Alternative inpainting techniques were  proposed in the past, through sparse recovery techniques \citep{starck:pires08,lanusse16} or Gaussian constraint realizations \citep{zaroubi95,starck:jeffrey18}. Since the Wiener model assumed the solution to be a Gaussian random field, it would make sense  to have a solution where the inpainted area presents the same statistical properties as in non inpainted aera. This property is by construction verified with constraint realizations, and it was shown it is also the case with sparse inpainting \citep{starck:pires08}. 
A similar sparse inpainting can be very easily included in the proximal Wiener filtering, minimizing the following equation:
\begin{align}
%%  \underset{\boldsymbol \kappa_{G}}{\mathrm{min}} 
\kw & = \argmin_{\kb } \left\{  \| \boldsymbol M(\gamma  - \A \kb ) \|^2_{\covn} + \beta \| \kb \|^2_{\covk}  + \lambda  \| \boldsymbol  \Phi^{*} \kb  \|_p \right\},
 \label{eq:optw_inp}
\end{align}
 where $p=0$ or $1$, $\Phi$ is the Discrete Cosine Dictionary (DCT), and $\lambda$ is a Lagrangian parameter. 
We end up with the forward-backward algorithm, called {\bf InpWiener}:
\begin{itemize}
\item Forward step:  
\begin{align}
\mathbf{t} = {\kb}^n   + 2 \mu \At {\cov}_{n}^{-1} (M(\gb - \A  {\kb}^n))
\label{eq:alg1_InpWiener}
\end{align}
\item Backward step:
\begin{align}
{\kb}^{n+1} = M    \boldsymbol u + (1-M)     {\boldsymbol \Delta}_{{\Phi}, \lambda} \boldsymbol u
\end{align}
where and ${\boldsymbol \Delta}_{{\Phi}, \lambda}$ is the proximal operator defined in  \citet{starck:pires08} which consists  in applying a DCT transform to $u$, threshold the DCT coefficients  and reconstruct an image from the thresholded coefficients, and 
\begin{align}
\boldsymbol u = \boldsymbol F^{*} \left( \mathbf{P}_{\kb}  \left( \mathbf{P}_\eta +  \mathbf{P}_{\kb} \right)^{-1} \right) \boldsymbol F \mathbf{t},
 \label{eq:alg2_InpWiener}
 \end{align}
\end{itemize}
 Areas where we have information are processed as in the usual Wiener case, while the inpainting regularization impacts area with missing data (i.e. when  $M=0$).
 
 Concerning Eq.~\ref{eq:optw_inp}, it is interesting to note that:
 \begin{itemize}
      \item{\bf KS:} if $\beta=0,\lambda=0$ and $\covn$ is diagonal with constant values along the diagonal (i.e. stationary 
      Gaussian noise),  {\bf InpWiener} leads to the non-iterative standard Kaiser-Squires solution. 
      \item{\bf GKS:}  If $\beta=0$ and $\lambda=0$,
      the least square estimator is derived with the iterative algorithm: ${\kb}^{n+1} = {\kb}^n   + 2 \mu \At {\cov}_{n}^{-1} (M(\gb - \A  {\kb}^n))$, with $\mu = \min (\covn) $. 
      As it generalizes the Kaiser-Squires method, we will call this algorithm {\bf GKS}.
     \item{\bf FASTLens:}  If $\beta=0$ and $\covn$ is diagonal with constant values along the diagonal, 
      {\bf InpWiener} leads to the FASTLens inpainting algorithm \citep{starck:pires08}.
      \item{\bf GIKS:} If $\beta=0$,  {\bf InpWiener} leads to an inpainted generalized the Kaiser-Squires
     solution where the {\bf InpWiener} Forward is unchanged, and the Backward step becomes:
     \begin{align}
{\kb}^{n+1} = M    \boldsymbol t + (1-M)     {\boldsymbol \Delta}_{{\Phi}, \lambda} \boldsymbol t
\end{align}
 \end{itemize}
Similarly to sections~\ref{sect_bmode} and \ref{sect_bmode}, these algorithms can handle data on the sphere and reconstruct jointly E and B modes.

\subsubsection*{Inpainted Wiener Experiment}
To test {\bf InpWiener}, we use the public MICE (v2) simulated galaxy catalogue presented Appendix~A.

\begin{figure*}
\centerline{
\hbox{
        \includegraphics[width=0.35\textwidth]{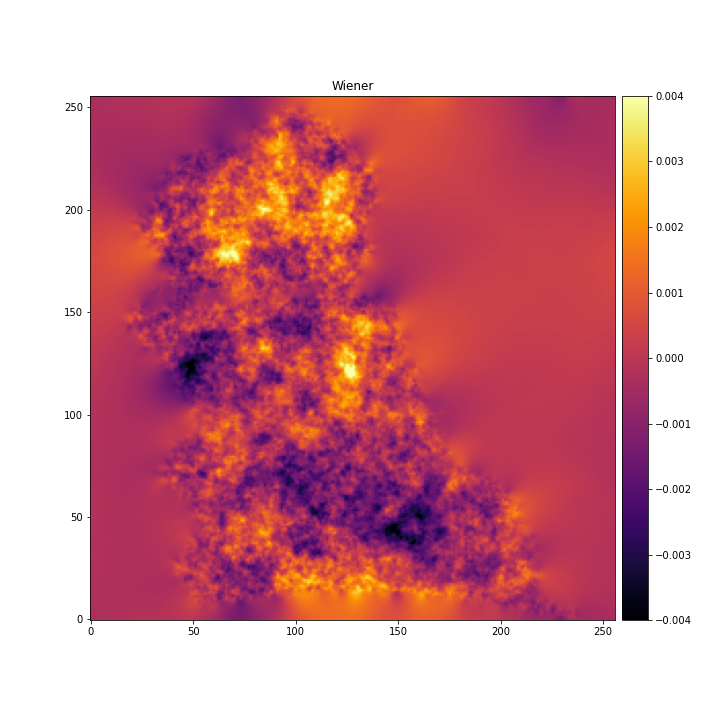}
        \includegraphics[width=0.35\textwidth]{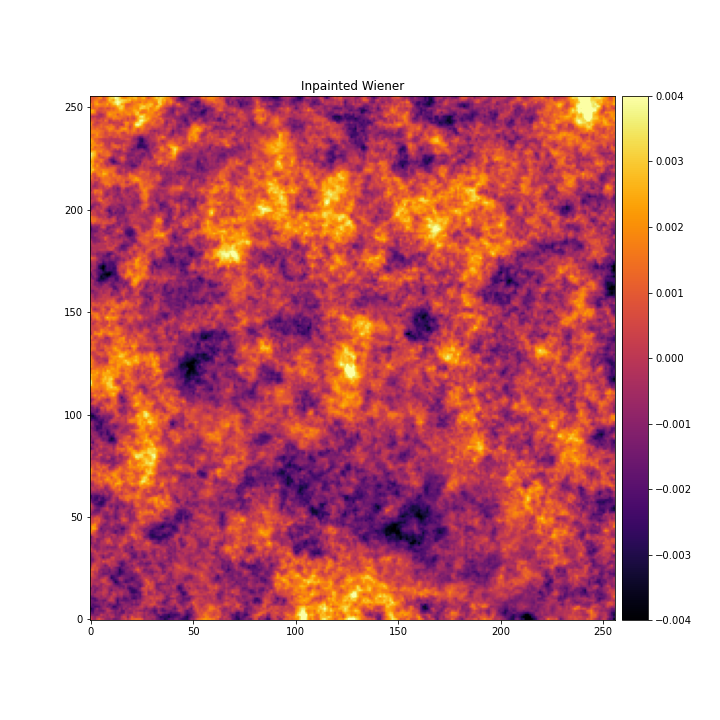}
}}
        \caption{Wiener and inpainted Wiener solutions.  }
        \label{fig:inp}
\end{figure*}

Figure \ref{fig:inp} shows the Wiener solution (left) and the inpainted Wiener solution (right) derived from the shear components shown in Fig.~\ref{fig:mice_shear}.  

% Given the high level of noise, which is common in real observations, correlations between the true convergence and the shear field are not detectable by eye.

% ===========================

\subsection{Agnostic Wiener Filtering}

The Wiener method needs to know the theoretical power spectrum $P_\kappa$, and the solution therefore varies with the assumed cosmological model used
to derive $P_\kappa$. To avoid this issue, a solution could to estimate $P_\kappa$ directly from the shear measurements, for instance  
using a mask correction as in \citet{lee2020}. Bayesian techniques have also been used to infer both the map and power spectrum \citep{Wandelt2004,jash2014,Alsing2016}. An alternative approach is to used the {\bf GIKS} inpainting algorithm to fill first the  missing area and  then to compute
the power spectrum of the inpainted map. 
% This can be also be achieved using the  Forward-Backward proximal algorithm, where the first  is the same as for the Wiener method, i.e. a gradient step, and the second step is replaced by prox function related to a sparse penalization as as described above. 
% This Inpainted Least Square convergence map estimator (ILS)  leads to the following iteration:
% \begin{align}
%{\kb}^{n+1} =  {\boldsymbol \Delta}_{{\Phi}, \lambda_n}  {\kb}^n   + 2 \mu \At {\cov}_{n}^{-1} (\gb - \A  %{\kb}^n)
%\end{align}
%with $\lambda_n$ decaying to zero with the iteration. 
Applying {\bf GIKS} on the data and a set of $R$ noise realizations, the final estimator is 
\begin{align}
P_{\kappa}  =  \mathrm{powspec} ( {\kb}_{Data} )  - \frac{1}{R}   \sum_i   \mathrm{powspec} ({\kb}_{Rea_i} ).
\end{align}
Since the data noise $P_{\kappa}$ will still be noisy, a final denoising step or a function fitting can be done.
\begin{figure}
\centerline{
\hbox{
        \includegraphics[width=.45\textwidth]{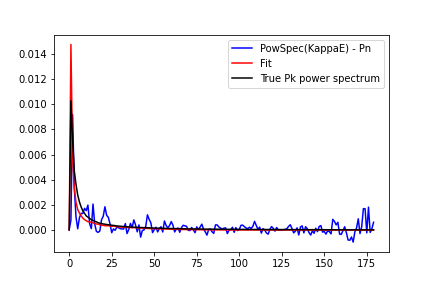}
}}
        \caption{Theoretical power spectrum of one convergence map. In black the true theoretical power spectrum; in red, the spectrum estimated using the {\bf GIKS} algorithm, corrected from noise power spectrum, and in blue the fit.}
        \label{fig:mice_pk}
\end{figure}

As an illustration, we fitted 
the function $f(k, a,u,e,c) = \exp(\mid a * (u k)^{-e}  \mid ) +  c$ to the estimated noisy $P_{\kappa} $.
Figure \ref{fig:mice_pk} shows an example of an estimated power spectrum following this approach.

\subsubsection*{Experiment: Impact of an unknown theoretical power spectrum}

In this experiment, we used the same public MICE simulations, and we extracted  18 shear different shear maps with different noise realizations.
For each of them, we applied  the forward-backward Wiener algorithm  using the true theoretical power spectrum, and we applied the inpainted agnostic Wiener method  on the same data, re-estimating for each of the 18 shear data sets the theoretical power spectrum. 
\begin{figure}
        \centering
        \includegraphics[width=.5\textwidth]{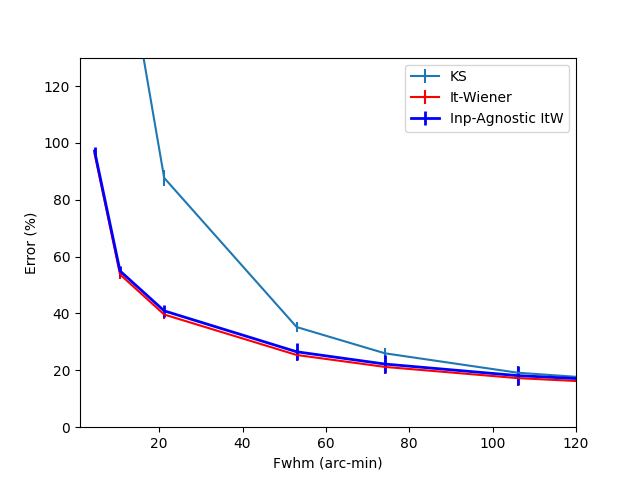}
        \caption{Reconstruction error at different resolution for Kaiser-Squires, Wiener and inpainted agnostic Wiener.  }
        \label{fig:agnostic}
  \end{figure}
Fig.~\ref{fig:agnostic} shows the reconstruction errors at different resolutions,  for Kaiser-Squires, Wiener and inpainted agnostic Wiener. 
We can easily see that the inpainting has no impact on the reconstruction error, which is expected since only the area where the mask is equal to one is used
in the error calculation,  and also that the agnostic approach also has very little impact on the final solution.  We do not claim that we should use an agnostic approach when applying a Wiener filtering, but 
it is interesting to have this option available, i.e. when using Wiener without any assumption about the cosmology, and to give us the possibility to check if cosmological priors impact the results. 

\end{appendix}

\bibliographystyle{bibtex/aa}
\bibliography{bibtex/refs,bibtex/JLSBibTex}

\end{document}